\documentclass[fleqn,usenatbib]{mnras}
\usepackage{dcolumn}
\usepackage{graphicx}
\usepackage{url}
\usepackage{color}
\usepackage[T1]{fontenc}
\usepackage{ae,aecompl}
\usepackage{pdflscape}
\usepackage[normalem]{ulem}
\usepackage{array}
\usepackage{epsf}

\usepackage{units}

\usepackage[varg]{txfonts}

\usepackage{verbatim}

\newcommand{\cm}{cm$^{-1}$}

\newcommand{\ai}{\textit{ab initio}}
\newcommand{\Ai}{\textit{Ab initio}}

\newcommand{\Duo}{{\sc Duo}}

\newcommand{\NO}[2]{$^{#1}$N$^{#2}$O}

\newcommand{\pp}{^{\prime\prime}}

\title[ExoMol Line Lists XXI:  NO]{ExoMol Line List XXI: Nitric Oxide (NO)}

\author[Andy Wong et al.]{ Andy Wong$^{1}$, Sergei N. Yurchenko$^{2}$, Peter Bernath$^{1}$,
 Holger S. P. M\"uller$^{3}$, \newauthor   Stephanie McConkey$^{2}$, Jonathan Tennyson$^{2}$\thanks{Email: j.tennyson@ucl.ac.uk}
\\
$^{1}$ Department of Chemistry and Biochemistry, Old Dominion University, 4541 Hampton Boulevard, Norfolk, VA 23529, USA \\
$^{2}$Department of Physics and Astronomy, University College London, London WC1E 6BT, UK \\
$^{3}$I.~Physikalisches Institut, Universit{\"a}t zu K{\"o}ln, Z{\"u}lpicher Str. 77, 50937 K{\"o}ln, Germany
}

\begin{document}

\label{firstpage}
\pagerange{\pageref{firstpage}--\pageref{lastpage}}

\maketitle

\begin{abstract}

Line lists for the X~$^2\Pi$ electronic ground state for the parent isotopologue of
nitric oxide ($^{14}$N$^{16}$O) and five other major isotopologues
($^{14}$N$^{17}$O, $^{14}$N$^{18}$O, $^{15}$N$^{16}$O, $^{15}$N$^{17}$O and
$^{15}$N$^{18}$O) are presented. The line lists are constructed using empirical energy levels (and line positions) and high-level \textit{ab inito} intensities. The
energy levels were obtained using a combination of two approaches, from an effective
Hamiltonian and from solving the rovibronic Schr\"{o}dinger equation variationally.
The effective hamiltonian model was obtained through a fit to the experimental line positions of NO available in the literature for all six isotopologues using the programs \textsc{SPFIT} and \textsc{SPCAT}. The variational model was built through a least squares fit of the \textit{ab inito}  potential and spin-orbit curves to the experimentally derived energies and experimental line positions of the main isotopologue only using the \textsc{Duo} program. The \textit{ab inito} potential energy, spin-orbit and dipole moment curves (PEC, SOC and DMC) are computed using high-level \textit{ab inito}  methods and the \textsc{MARVEL} method is used to obtain energies of NO from experimental
transition frequencies.
The line lists are constructed for each isotopologue based on the use of the most accurate
energy levels and the \textit{ab inito} DMC. Each line list covers a wavenumber
range  from 0 - 40,000 \cm with approximately 22,000 rovibronic states
and 2.3 -- 2.6 million transitions extending to $J_{max} = 184.5$ and $\varv_{max} = 51$.
Partition functions are also calculated up to a temperature of 5000~K.
The calculated absorption line intensities at 296~K using these line lists
show excellent agreement with those included in the
HITRAN and HITEMP databases. The computed NO line lists are the most comprehensive to date, covering a wider wavenumber and temperature range
compared to both the HITRAN and HITEMP databases.
These line lists are also more accurate than those used in HITEMP.
The full line lists are available from the CDS \url{http://cdsarc.u-strasbg.fr} and ExoMol
 \url{www.exomol.com} databases; data will also be available from CDMS \url{www.cdms.de}.

\end{abstract}


\section{Introduction} \label{intro}

NO has been detected in several interstellar environments ranging from a
starburst galaxy \citep{03MaMaMa.NO, 06MaMaMa.NO} to dark clouds
\citep{90McZiIr.NO} and numerous star-forming regions
\citep{91ZiMcMi.NO}. It is also present in the atmospheres of Earth, Mars
and Venus, and its emission is a major source of nightglow in these three
planets \citep{08CoSaGe.NO, 92EaHuLe.NO, 10RoMoBe.NO}. The presence of
NO in Earth's atmosphere has a significant impact on
depletion of the ozone layer \citep{10BaChxx.book, 00Waxxxx.book} and originates from the
reaction of N$_2$O with O($^1$D) in the stratosphere. In the troposphere
the major sources of NO are of anthropogenic origin as it is produced during fuel combustion
at temperatures of $\sim$2300~K \citep{88FlSexx.book} and through soil
cultivation. Because NO (and NO$_2$) catalyze the production of tropospheric ozone,
it is important to try to reduce NO emissions (amongst other air pollutants).
Whilst NO has not yet been detected in an exoplanet atmosphere, it
is likely present in its gaseous form in terrestrial-type
atmospheres, for example produced during a storm soon after a lightning shock \citep{17ArRiWa}.

There are currently two commonly used databases which contain line lists for NO in its electronic ground state:
 HITRAN \citep{HITRAN2012} which covers
$^{14}$N$^{16}$O, $^{14}$N$^{18}$O and $^{15}$N$^{16}$O,
and is designed for use near room temperature,
and HITEMP \citep{jt480} which allows the spectrum
of $^{14}$N$^{16}$O to be modelled up
to 4000~K. HITRAN and HITEMP databases contain 103,702
and 115,610 transitions, respectively,
with ${J_{max}=125.5}$ and ${v_{max}=14}$.

The aim of this work is to produce a line list for the X~$^2\Pi$ electronic
ground state of nitric oxide for its parent isotopologue ($^{14}$N$^{16}$O, hereafter NO), and
five major isotopologues ($^{14}$N$^{17}$O,
$^{14}$N$^{18}$O, $^{15}$N$^{16}$O,
$^{15}$N$^{17}$O and $^{15}$N$^{18}$O). Using a combination of high level \ai\
methods, accurate fitting to a comprehensive set of experimental data and variational modeling,
six line lists for these species were constructed. These line lists
consist of rovibronic energy levels, with all of the associated quantum numbers, transition wavenumbers and
Einstein-A coefficients.  The computed line lists form
part of the ExoMol database \citep{jt631} which aims to
provide a comprehensive set of high-temperature line lists for
molecules that may be present in hot atmospheres such as those of
exoplanets, planetary disks, brown dwarfs and cool stars
\citep{12TeYuxx.db}.
Many ExoMol line lists have already been used in the
characterisation and modelling of brown dwarf and exoplanet
atmospheres
\citep{jt400,11CuKiDa.dwarfs,jt572,15MoBoDu.exo,15MoFoMa.exo,15BaKoQu.exo,11BeTiKi.exo,15CaLuYu.dwarfs,14MoMaFo.dwarfs,jt629}.

The ExoMol database also contains line lists for numerous diatomic
molecules generated by the ExoMol project: AlO \citep{jt598}, BeH, MgH and CaH
\citep{jt529}, SiO \citep{13BaYuTe.SiO}, CaO \citep{16YuBlAs.CaO}, CS
\citep{15PaBaYu.CS},  NaCl and KCl
\citep{14BaChGo.NaCl}, NaH \citep{15RiLoYu.NaH}, PN
\citep{jt590}, ScH \citep{15LoYuTe.ScH} and VO \citep{jt644}.
Line lists available from other sources include: CrH
\citep{02BuRaBe.CrH}, FeH \citep{03DuBaBu.FeH}, TiH
\citep{05BuDuBa.TiH}, CaH \citep{11LiHaRa.cah}, MgH
\citep{2013MgH-MNRAS}, CN \citep{14BrRaWe.CN}, OH \citep{2016OH-JQSRT}, NH \citep{2015NH-JCP}, and ZrS \citep{17Farhat.ZrS}.

These line lists often include many of the abundant isotopologues and greatly
extend the calculated range of ${J}$ and ${v}$
in comparison to the HITRAN and HITEMP databases.
 High-temperature line lists are useful in the characterisation of brown dwarfs \citep{jt572}, which have atmospheric
temperatures ranging from 500-3000~K \citep{14Perryman.book}.
The hot NO line lists produced from this work can be used directly in characterization
of the spectra of such objects, as well as in atmospheric
models \citep{13VeSexx.book}.

The remainder of this paper is divided into several sections. Section \ref{methods}
outlines the methods used in the calculation of energy levels and
production of the line lists, whilst Section \ref{results} presents
the results of this work - mainly the NO line list, the calculated partition function
and radiative lifetimes.
Absorption line intensities and cross-sections are
also shown.  Finally, Section \ref{conc} discusses the
implications of this work, and how it will be significant to the
astrophysical community.

\section{Methods}
\label{methods}

\subsection{Experimental Data}
\label{expt_data}

\subsubsection{Extraction of Experimental Data}	
\label{expt_data2}

Transition frequencies for the X~$^2\Pi$ electronic ground state of NO were collected from
selected experimental papers, see Table~\ref{table:expt_papers}, along
with any given quantum numbers and uncertainties for all six isotopologues. This table also indicates whether
a dataset was used for the \textsc{MARVEL} analysis (M),  \textsc{SPFIT} calculations (S) or both (see below).
We used the \textsc{MARVEL} program (Measured Active
Rotational-Vibrational Energy Levels)
\citep{MARVEL,12FuCsi.method} to derive
energy levels of the main isotopologue \NO{14}{16}\
based on the experimental transition data available in the literature. These data were then utilized to refine our
\ai\ model using the \Duo\ program \citep{Duo}. The more extensive \textsc{SPFIT}  set of experimental frequencies covering all
six isotopologues was used to obtain NO spectroscopic constants in a global fit using the effective
Hamiltonian programs \textsc{SPFIT} and \textsc{SPCAT} \citep{spfit_1991}.


\begin{table*}
\caption{Experimental papers on NO spectra. }
\label{table:expt_papers}
\begin{tabular}{llllllll}
\hline
\hline
Source                            &  States                            & Isotopologue; Methods$^a$                    & M/S$^b$          & Range in $J$ and/or $\varv$  \\
\hline
{\footnotesize 64James}           &    \citet{64James.NO}              &    NO; IR                                                 &   M       &    $J=0.5-21.5$                                       \\
{\footnotesize 64JaThxx}          &    \citet{64JaThxx.NO}             &    NO; IR                                                 &   M       &    $J=1.5-21.5$	                                      \\
{\footnotesize 70Neumann}         &    \citet{70Neumann.NO}            &    NO; RF, $\mu$                                          &   S       &    $J=0.5-7.5$, $\varv = 0$                           \\
{\footnotesize 72MeDyxx}          &    \citet{72MeDyxx.NO}             &    NO, $^{15}$NO; IR                                      &   S       &    $J=0.5-8.5$, $\varv = 0$                           \\
{\footnotesize 76Meerts}          &    \citet{76Meerts.NO}             &    NO; RF                                                 &   S       &    $J=4.5-5.5$, $\varv = 0$                           \\
{\footnotesize 77DaJoMc}          &    \citet{77DaJoMc.NO}             &    NO; IR-RF DR                                           &   S       &    $J=4.5-5.5$, $\varv = 0$, 1                        \\
{\footnotesize 78AmBaGu}          &    \citet{NO_v1-0_2-1_plus_1978}   &    NO; FTIR                                               &   S       &    $J=0.5-40.5$, $\varv = 1-0$, $2-1$                 \\
{\footnotesize 78HeLeCa}          &    \citet{78HeLeCa.NO}             &    NO; FTIR                                               &   S       &    $J=0.5-28.5$, $\varv = 3 - 0$                      \\
{\footnotesize 79AmGuxx}          &    \citet{565758_IR_1-0_+_1979}    &    $^{15}$NO, $^{15}$N$^{17}$O, $^{15}$N$^{18}$O; FTIR    &   S       &    $J=0.5-41.5$, $\varv = 1-0$, $2-1$                 \\
{\footnotesize 79PiCoWa}          &    \citet{79PiCoWa.NO}             &    NO; mmW, smmW                                          &   S       &    $J=0.5-4.5$, $\varv = 0$                           \\
{\footnotesize 80AmVexx}          &    \citet{80AmVexx.NO}             &    NO; Emi, FTIR 2900$-$3810 \cm.                         &   M,S     &    $J=0.5-57.5$, $\varv=0-15$, $\Delta v=2$           \\
{\footnotesize 80TeHeCa}          &    \citet{80TeHeCa.NO}             &    $^{15}$NO, $^{15}$N$^{18}$O; FTIR                      &   S       &    $J=0.5-32.5$, $\varv = 1-0$, $2-0$, $3-0$          \\
{\footnotesize 80VaMeDy}          &    \citet{80VaMeDy.NO}             &    NO; TuFIR                                              &   M       &    $J=7.5-10.5$                                   \\
{\footnotesize 81LoMcVe}          &    \citet{81LoMcVe.NO}             &    NO; IR-RF DR                                           &   S       &    $J=12.5-20.5$, $\varv = 0$, 1                      \\
{\footnotesize 82Amiot}           &    \citet{82Amiot.NO}              &    NO; Emi, FTIR, 3800$-$5000 \cm                         &   M,S     &    $J=0.5-59.5$, $\varv=7-22$, $\Delta \varv=3$       \\
{\footnotesize 86HiWeMa}          &    \citet{86HiWeMa.NO}             &    NO; heterodyne IR                                      &   S       &    $J=1.5-32.5$, $\varv = 1 - 0$                      \\
{\footnotesize 91SaYaWi}          &    \citet{91SaYaWi.NO}             &    $^{15}$NO, N$^{18}$O$^c$; mmW, smmW                    &   S       &    $J=0.5-4.5$, $\varv = 0$                           \\
{\footnotesize 92RaFrMi}          &    \citet{92RaFrMi.NO}             &    NO; IR ChLumi, 2.7$-$3.3${\mu}$m \                     &   M       &    5.2$-$6.8 $\mu$m $\varv'=2-3$                     \\
{\footnotesize 94DaMaCo}          &    \citet{94DaMaCo.NO}             &    NO; FTIR                                               &   M,S     &    $J=2.5-24.5$, $\varv = 2 - 1$                      \\
{\footnotesize 94MaDaCo}          &    \citet{94MaDaCo.NO}             &    NO; FTIR                                               &   M       &    $J=1.5-20.5$, $\varv = 1 - 0$                      \\
{\footnotesize 94SaLiDo}          &    \citet{94SaLiDo.NO}             &    N$^{17}$O and $^{15}$N$^{18}$O; mmW                    &   S       &    $J=0.5-2.5$, $\varv = 0$                           \\
{\footnotesize 94SpChGi}          &    \citet{94SpChGi.NO}             &    NO; FTIR 1780$-$1960 \cm                               &   M,S     &    $J=0.5-25.5$, $\varv = 1 - 0$                      \\
{\footnotesize 95CoDaMa}          &    \citet{95CoDaMa.NO}             &    NO; FTIR$^d$ 1730$-$1990 \cm                           &   M,S     &    $J=0.5-41.5$, $\varv = 1 - 0$                      \\
{\footnotesize 96SaMeWa}          &    \citet{96SaMeWa.NO}             &    NO; heterodyne IR                                      &   S       &    $J=8.5-18.5$, $\varv = 1 - 0$                      \\
{\footnotesize 97DaDoKe}          &    \citet{97DaDoKe.NO}             &    O$_2$/N$_2$; Ebert spectrograph                       &   M       &    $\varv=0-7$                                        \\
{\footnotesize 97MaDaRe}          &    \citet{97MaDaRe.NO}             &    NO; FTIR, 3600$-$3800 \cm                              &   M,S     &    $J=2.5-32.5$, $\varv = 2 - 0$                      \\
{\footnotesize 98MaDaRe}          &    \citet{98MaDaRe.NO}             &    NO; FTIR, 3600$-$3720 \cm                              &   M       &    $J=2.5-17.5$, $\varv = 3 - 1$                      \\
{\footnotesize 99VaStEv}          &    \citet{99VaStEv.NO}             &    NO; TuFIR, 11$-$157 \cm                     &   M       &    $J=2.5-38.5$                                       \\
{\footnotesize 99VaStEv}          &    \citet{99VaStEv.NO}             &    NO; $^{15}$NO; TuFIR, 11$-$157 \cm                     &   S       &    $J=2.5-38.5$, $\varv = 0$                          \\
{\footnotesize 01LiGuLi}          &    \citet{NO_v01_dip_2001}         &    NO; IR LMR, $\mu$                                      &   S       &    $J=1.5-2.5$, $\varv = 1 - 0$                       \\
{\footnotesize 06LeChOg}          &    \citet{06LeChOg.NO}             &    NO; FTIR                                               &   M       &    $J=0.5-30.5$, $\varv'=2-6$	                    \\
{\footnotesize 06BoMcOs}          &    \citet{06BoMcOs.NO}             &    NO; NICE-OHMS                                          &   M       &    $J=2.5-16.5$, $\varv = 7 - 0$                      \\
{\footnotesize 15MuKoTa}          &    \citet{15MuKoTa.NO}             &    N$^{18}$O; TuFIR, 33$-$159 \cm                         &   S       &    $J=3.5-26.5$, $\varv = 0$                          \\
\hline\hline
\end{tabular}
{\\
$^a$ Unlabelled atoms refer to $^{14}$N or $^{16}$O.
Abbreviations: IR (infrared), FT (Fourier transform), ChLumi (chemiluminescence), Emi (emission), DR (double resonance), $\mu$ (dipole moment),
RF (radio frequency), MW (microwave), mmW (millimetre wave), smmW (sub-millimetre wave), TuFIR (tunable far-infrared), LMR (laser magnetic resonance).\\
$^b$ Used for  \textsc{MARVEL} (M) and/or  \textsc{SPFIT} (S); \\
$^c$ NO FIR data not used. $^d$ Wavenumber recalibration proposed, see section~\ref{section_NO_spec-parameters}.}
\end{table*}


\subsubsection{MARVEL}	\label{marv}

\textsc{MARVEL} is an algorithm
which calculates rovibronic energy levels from a given set of
experimental transitions.
The online\footnote{http://kkrk.chem.elte.hu/Marvelonline} version of the
program was used, as the output files are formatted to be more
user-friendly. An extraction from our \textsc{MARVEL} input-file is given in
Table~\ref{t:MARVEL:format}.

The experimental literature was chosen to ensure that a full range of transitions was included whilst minimising duplication of data.
Although the default {\sc MARVEL} procedure utilises all of the data included in the input file \citep{12FuCsxx.methods},
if there happens to be any data overlap between papers, only the
most  accurate (and in general the most recent) data are considered. Only transitions within the ground electronic state were considered for our {\sc MARVEL} analysis.

Pure rotational transition frequencies were taken from
\citet{80VaMeDy.NO}, \citet{74LoTixx.NO} and \citet{99VaStEv.NO}, whilst transitions between the two
$\Lambda$-doublet states (X~$^2\Pi_{\nicefrac{1}{2}}$ and X~$^2\Pi_{\nicefrac{3}{2}}$) were taken from
\citet{94MaDaCo.NO}. Rovibrational transitions, extending up to
$\Delta \varv=3$, $\varv'=22$ and $J'=58.5$, were taken from the work by
\citet{82Amiot.NO}, \citet{80AmVexx.NO}, \citet{06BoMcOs.NO},
\citet{95CoDaMa.NO}, \citet{06LeChOg.NO}, \citet{97MaDaRe.NO},
\citet{98MaDaRe.NO} and \citet{94SpChGi.NO}.

It should be noted that the spectroscopic notation is not consistent in the literature, thus making it necessary to
generate a consistent set of quantum numbers for each transition. For
many of the rovibrational papers, the $P$, $Q$ and $R$ labels were used to
derive $J'$ if $J'$ and $J\pp$ had not already been specified.
In the case of \citet{82Amiot.NO} and \citet{80AmVexx.NO}, the projection of the total angular momentum ($\Omega$)
was determined by assigning transitions labelled $P_1$ and $R_1$ to a value of
$\Omega\pp= \frac{1}{2}$ and those labelled $P_2$ and $R_2$ to the corresponding $\Omega\pp=\frac{3}{2}$.
Rotationless parities of lower levels were given in terms of $e$ and $f$ by
\citet{95CoDaMa.NO}, \citet{94MaDaCo.NO}, \citet{97MaDaRe.NO} and
\citet{94SpChGi.NO}. For papers that did not specify parity, this was resolved by duplicating the dataset
and assigning the parity $e$ to one set and the parity $f$ to the other.
Hyperfine splitting was also resolved, albeit only in a few pure rotational papers
(\citet{80VaMeDy.NO}, \citet{74LoTixx.NO} and \citet{99VaStEv.NO}), however at this stage of our analysis the hyperfine splitting was ignored.
For rovibrational transitions sharing the same quantum numbers ($\Omega$', $\Omega\pp$, $J'$, $J\pp$ and
$e/f$) but different frequency, an average of the two frequencies was taken and the frequency uncertainty
was propagated. In lieu of any specified transition frequency uncertainties,
estimates were made based upon the precision with which frequencies
were quoted.


\begin{table*}
\caption{{\sc MARVEL} format of the experimental data: extract
from the MARVEL input file.}
\makebox[\textwidth]{%
\small
 \begin{tabular}{rrrcccrcccl}
 \hline\hline
 Wavenumber (\cm) & Uncertainty (\cm) & $J'$ & Parity' & $\varv'$ & $\Omega$'  & $J\pp$ & Parity'' & $\varv\pp$ & $\Omega\pp$  & Reference \\
 \hline
  1985.3307  &   0.005  &    42.5  &   -   &  1   &    0.5   &     41.5  &   +   &      0   &    0.5   &    95CoDaMa80 \\
  1806.6561  &   0.005  &    17.5  &   +   &  1   &    1.5   &     18.5  &   -   &      0   &    1.5   &    95CoDaMa81 \\
  1806.6561  &   0.005  &    17.5  &   -   &  1   &    1.5   &     18.5  &   +   &      0   &    1.5   &    95CoDaMa82 \\
  1802.5924  &   0.005  &    18.5  &   +   &  1   &    1.5   &     19.5  &   -   &      0   &    1.5   &    95CoDaMa83 \\
  1802.5924  &   0.005  &    18.5  &   -   &  1   &    1.5   &     19.5  &   +   &      0   &    1.5   &    95CoDaMa84 \\
  1798.4950  &   0.005  &    19.5  &   +   &  1   &    1.5   &     20.5  &   -   &      0   &    1.5   &    95CoDaMa85 \\
  1798.4950  &   0.005  &    19.5  &   -   &  1   &    1.5   &     20.5  &   +   &      0   &    1.5   &    95CoDaMa86 \\
  1794.3687  &   0.005  &    20.5  &   +   &  1   &    1.5   &     21.5  &   -   &      0   &    1.5   &    95CoDaMa87 \\
  1794.3687  &   0.005  &    20.5  &   -   &  1   &    1.5   &     21.5  &   +   &      0   &    1.5   &    95CoDaMa88 \\
  1790.2060  &   0.005  &    21.5  &   +   &  1   &    1.5   &     22.5  &   -   &      0   &    1.5   &    95CoDaMa89 \\
  1790.2060  &   0.005  &    21.5  &   -   &  1   &    1.5   &     22.5  &   +   &      0   &    1.5   &    95CoDaMa90 \\
  1786.0180  &   0.005  &    22.5  &   +   &  1   &    1.5   &     23.5  &   -   &      0   &    1.5   &    95CoDaMa91 \\
 \hline\hline
\end{tabular}}
\label{t:MARVEL:format}
\end{table*}

The $e/f$ parity of the lower energy states were converted to +/-
total parity using the following standard relations:
$$
\begin{array}{cl}
e: & {\rm parity} = (-1)^{J-\frac{1}{2}}\\
f: & {\rm parity} = (-1)^{J+\frac{1}{2}},
\end{array}
$$
where $-1$ and $+1$ corresponds an odd (--) parity and even parity (+), respectively.
The + $\leftrightarrow$ -- selection rule was used to determine the parity of the upper state.
For each transition, the electronic state and the projection of the
electronic angular $\Lambda$ remained unchanged as
only ro-vibrational transitions within the X~$^2\Pi$ electronic ground state
were considered in this work.

The quoted uncertainty of some transition frequencies were found to be either too large or too small.
In these cases, the uncertainty value was adjusted to agree
with the \textsc{MARVEL}-suggested uncertainty.
After some trial and error, a clean run in \textsc{MARVEL} with no errors was
achieved yielding a network of 4106 energy levels for NO with
$\varv_{\rm max} = 22$, $J_{\rm max} = 58.5$ and a term value maximum of 36,200~\cm.
The \textsc{MARVEL} energies obtained and the input file containing experimental transition frequencies are given as supplementary material to this paper.

The MARVEL procedure has previously been used to treat two other
open shell diatomics of astronomical importance: C$_2$ \citep{jt637}
and TiO \citep{jt672}. The treatment of a single,
albeit $^2\Pi$, state here proved to be much simpler than either of those
studies, which included a large number of electronic transitions.

\subsection{\textit{Ab Initio} calculations}	\label{Molpro}

\begin{figure}
\centering
  \centering
  \includegraphics[scale=0.3]{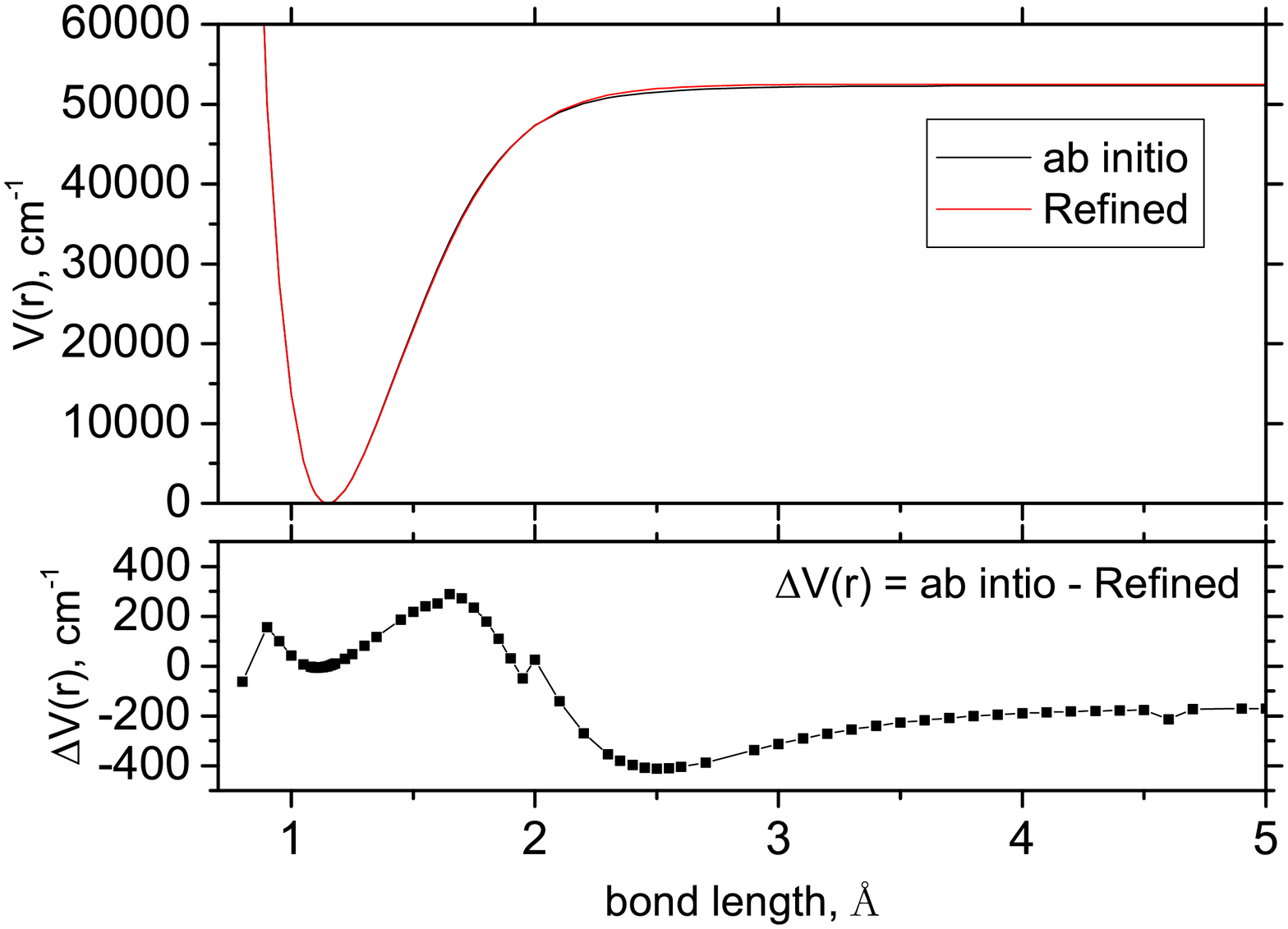}
  \centering
  \includegraphics[scale=0.3]{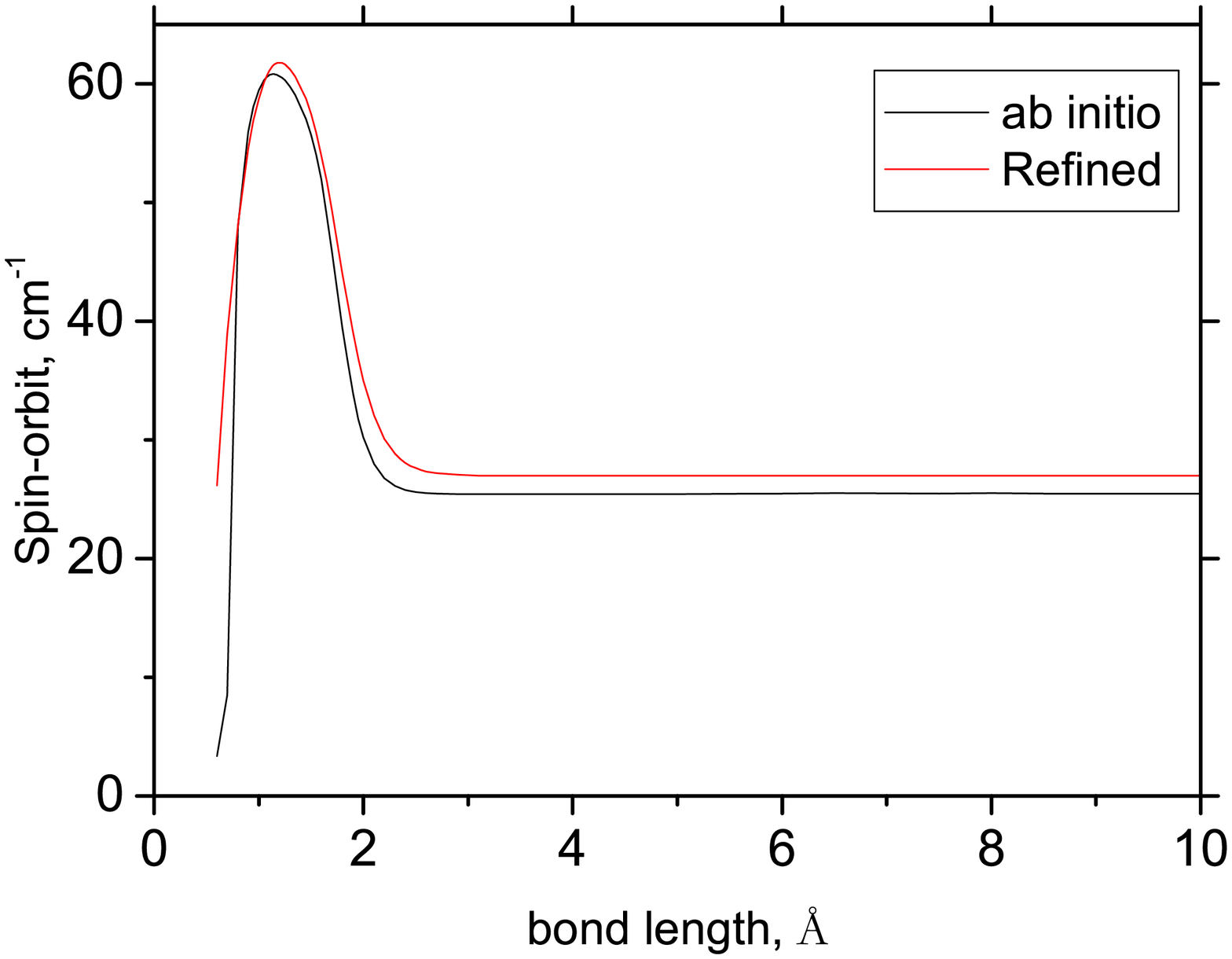}
\caption{\Ai\ and \textsc{Duo} refined PECs (upper display) and SOCs (lower display). The middle display shows the difference between the \ai\ and refined PECs.  }
\label{fig:curve_comp}
\end{figure}

The \ai\ potential energy curve (PEC), spin-orbit curve (SOC) and dipole moment curve (DMC) for the
X~$^2\Pi$ electronic ground state of NO were calculated using \textsc{MOLPRO}
\citep{MOLPRO}. An active space representation of (7,2,2,0) was chosen and
an internally contracted multireference configuration interaction (icMRCI) method was used
with Dunning type basis sets \citep{02PeDuxx.ai}. A quadruple-$\zeta$ aug-cc-pwCVQZ-DK basis set was
used to calculate the PEC and SOC, whereas
the DMC was calculated using a quintuple-$\zeta$ aug-cc-pwCV5Z-DK basis set. The range of 0.6 -- 10.0~\AA\ was used with a dense grid of 350 geometries. Relativistic corrections for the DMC were also evaluated based on the Douglas-Kroll-Hess (DKH) Hamiltonian which included core correlation.
The \ai\ PEC and SOC are shown in Fig.~\ref{fig:curve_comp}.

\begin{figure}
\centering
  \centering
  \includegraphics[scale=0.3]{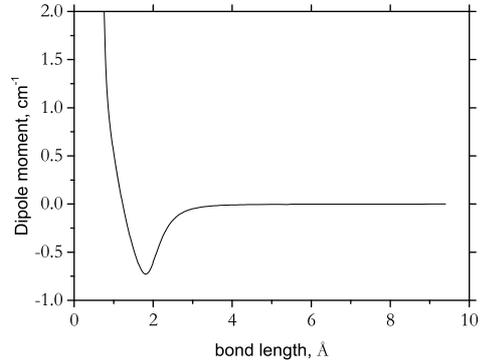}
\caption{\Ai\ MRCI/aug-cc-pwCV5Z-DK dipole moment curve of NO.  }
\label{f:DMC}
\end{figure}

A quadruple-$\zeta$ basis set was considered sufficient for the PEC
and SOC as these \ai\ curves  were fitted
to the experimental data using \Duo. A more accurate level of theory
MRCI/aug-cc-pwCV5Z-DK \cite{Werner1988,05BaPexx.ai,06BaPexx.ai} with the relativistic corrections based on the Douglas-Kroll-Hess hamiltonian and core-correlated was used for the DMC as implemented in MOLPRO \cite{MOLPRO}. The DMC was calculated using the energy-derivative method \citep{jt475} which calculates the dipole moment as a derivative of the electronic energy $E(F)$ with respect
to an external electric field $F$ ($F = 0.0005$~a.u. in this case)
using finite differences.
A dipole moment of ${\mu_e} = 0.166$~D at an equilibrium
internuclear distance $r_{\rm e} = 1.15~\AA\ $ was obtained.
\citet{70Neumann.NO} reported an experimental value for $\mu _0 = 0.15782~(2)$~D,
and \citet{NO_v01_dip_2001} determined $\mu _0 = 0.1595~(15)$~D and $\mu _1 = 0.1425~(16)$~D.
Our value agrees very well with the
value 0.1680~(19)~D for both $\mu _0$ and $\mu _1$ obtained using data from \citet{NO_v01_dip_2001}. The \ai\ DMC is shown in Fig.~\ref{f:DMC}.

\subsection{Duo: Fitting}	\label{Duo_fit}

\textsc{Duo} is a program designed to solve a coupled Schr{\"o}dinger
equation for the motion of nuclei of a given diatomic molecule
characterized by an arbitrary set of electronic states \citep{Duo}.
Based on Hund's case (a), \textsc{Duo} is capable of both refining potential energy curves (by
fitting data to experimental energies or transition frequencies) and producing line lists.
An extensive discussion of this method to calculate the direct solution of the vibronic Schr\"odinger
equation has been given in a recent topical review \citep{jt632}.

For this study, the range of computed $J$ levels was chosen to roughly correspond to all bound states of the system, i.e. to all states below $D_0$ ($J$ = 0.5 -- 184.5). The vibrational basis set was specified to have $\varv_{\rm max}$ = 51, which also corresponds to the maximal number of vibrational states (taken at $J=0$). The sinc DVR method \citep{Duo} defined on a grid of 701 points evenly distributed between 0.6 -- 4.0~\AA\ was used in integrations. The sinc DVR  allows one to reduce the number of points with no significant lost of accuracy. For example, all energies  obtained with this grid coincide with the energies obtained using a larger grid of 3001 points to better than $10^{-6}$~\cm.

The PEC and SOC were defined using the Extended Morse Oscillator (EMO)
potential \citep{EMO} given by
\begin{equation}
\label{eq:EMO}
V(r) = V_{\rm e} + D_{\rm e}\Bigg[1-\exp\Bigg(-\beta_{\rm EMO}(r)(r-r_{\rm e})\Bigg)\Bigg]^2 ,
\end{equation}
where $D_{\rm e}$ is the dissociation energy; $N$ is the expansion
order parameter;
$r_{\rm e}$ is the equilibrium internuclear bond distance;
$\beta_{\rm EMO}$ is the distance dependent exponent coefficient,
defined as
\begin{equation}
\label{eq:beta}
\beta_{\rm EMO}(r) = \sum\limits_{i=0}^{N} B_i y_{p} (r)^i
\end{equation}
and ${y_{p}}$ is the \u{S}urkus variable \citep{84SuRaBo.method} given by
\begin{equation}
\label{eq:y}
y_{p}(r) =  \frac{r^p - r^{p}_{\rm e}}{r^p + r^{p}_{\rm e}}
\end{equation}
with $p$ as a parameter.  The EMO form is our common choice for representing PECs of diatomics
\citep{jt598,16YuBlAs.CaO,15LoYuTe.ScH,jt644}. It guarantees the
correct dissociation limit and also allows extra flexibility in the degree
of the polynomial around a reference position $R_{\rm ref}$, which was defined as the
equilibrium internuclear separation ($r_{\rm e}$) in this case. It is also very robust in the fit.
The disadvantage of EMO is that it does not correctly describe the dissociative part of the curve.
As we show below,  this drawback does not have a significant impact on our line lists.

A reasonable alternative to EMO is the Morse/long-range (MLR) potential representation  \citep{07LeHexx.Ca2,09LeDaCo.Li2,11LeRoHa.MLR}, which guarantees a physically correct, multipole-type representation of the PEC as inverse powers of $r$ for $r\to \infty$. The disadvantage of MLR  (at least according to our experience) is that it less robust than EMO in refinements, requiring very careful determination of the switching and damping functions (see, for example, \citet{11LeRoHa.MLR}). Furthermore  \citet{jt442} showed that for strongly bound systems the multipole-type expansion is unnecessary and ``possibly harmful'', except for very large values of $r$ ($>10 a_0$ in case of H$_2$), which is certainly not the case here. Therefore our choice was to use EMO. As shown below, our line list is truncated at 40,000~\cm, and thus does not come very close to the long-range of the NO PEC.

The \ai\ PEC and SOC were fitted to the experimental line positions of \NO{14}{16}\ available in the literature combined with the experimentally derived energies generated by \textsc{MARVEL}. From our experience a combination of line positions and energy levels provides a more stable fit.
A total of nine potential expansion parameters ($B_0, \ldots ,B_8$) was required in order to obtain an optimal fit.
The addition of any more parameters did not improve the fit significantly.
In the case of the SOC refinement, an inverted EMO potential is used (Fig.~\ref{fig:curve_comp}) and required only four expansion parameters
($B_0, \ldots ,B_3$) to achieve a satisfactory fit.


The experimental value $D_0$ is 52,400~$\pm 10$~\cm\ estimated by \citet{70CaPixx}
using fluorescence experiments and from \citet{68AcMixx}.
We decided to refine the dissociation
energy ($D_{\rm e}$) and not to constrain it to the experimental value
of \citet{70CaPixx}.
Varying $D_{\rm e}$ parameter led to a more compact form of $\beta_{\rm EMO}(r)$  with $N=6$ instead $N=8$: less expansion parameters usually means a more stable extrapolation. In fact \citet{88DePexx.NO} noted in their MRD-CI study the change of the dominant character of the reference electronic configurations in NO PEC at about 3 and 5 bohr ($~$29100~\cm and $~51800$~\cm\ respectively), when approaching the dissociation  N($^4S$) + O($^3P$) (from $\pi^4 \pi^*$ to $\sigma \pi^3 \pi_x^* \pi_y^* \sigma^*$). That is, it is difficult to obtain a reliable connection between the equilibrium and experimental $D_0$ without sampling the highly vibrationally excited states ($v>48$) experimentally in the fit.
Due to the lack of these data and also that the dissociation region was not our priority, we decided to adopt the refined $D_{\rm e}$ value. Our final SO-free $D_0$ is 51608 (6.400~eV), which is of $~$800~\cm\ away from the experiment. This should not affect the quality of our  line list for the selected temperatures. The $D_0$ value is estimated using the refined value of
$D_{\rm e} = 52495.3$~\cm\  (see Table
\ref{table:P_S_params}) and the lowest energy relative to the PEC minimum $V_{\rm e}$ of
$E_{J=0.5, \Omega=0.5, v=0} = 887.100$~\cm. \citet{04PoFixx.NO} reported a  high-level \ai\ level $D_{\rm e}$ value (MR-ACPF(TQ)) of 51,140~\cm\ (6.340~eV).


The resulting EMO parameters, including the reference (equilibrium) bond length ($r_{\rm e}$), dissociation energy ($D_{\rm e}$), expansion ($B_n$) and $p$ parameters are listed in Table~\ref{table:P_S_params}.
The \ai\ potential energy and spin-orbit curves are compared to the
refined curves from \Duo\ in Fig.~\ref{fig:curve_comp}. The
\ai\ PEC is in good agreement with the refined PEC, despite the very
aggressive fit applied with a large number of $B_n$ expansion
parameters. The \ai\ spin-orbit curve was changed substantially by fitting, although the overall shape of the \ai\ SOC is maintained in the
refined curve, which is reassuring.

To account for spin-rotation and $\Lambda$-doubling effects, a polynomial expansion
based on the \u{S}urkus-variables was used:
\begin{equation}
\label{eq:bobleroy}
V(r) = D_{\rm e} + (1 - y^{eq}_p)\sum\limits^{N}_{n\geq 0} A_n (y^{eq}_p)^i
\end{equation}
where $N$ and $p$ are parameters, and $A_n$ is an expansion parameter refined in \textsc{Duo}. Both the spin-rotation $\gamma(r)$ and $\Lambda$-doubling $[p+2q](x)$ (see, for example \citet{79BrMexx.methods}) functions were fitted with two
expansion parameters $A_0$ and $A_1$. These are given in Table
\ref{table:sr_L_params} along with the $p$ and $N$ parameters.
Varying the Born-Oppenheimer breakdown corrections \citep{02LeHuxx.diatom} did not lead to a significant improvement, at least within the root-means squares achieved, and therefore were excluded from the fit. In fact the effective interaction with other electronic states was partly recovered by inclusion of the Lambda-doubling and spin-rotation effective functions.

\begin{figure}
\centering
\includegraphics[scale=0.3]{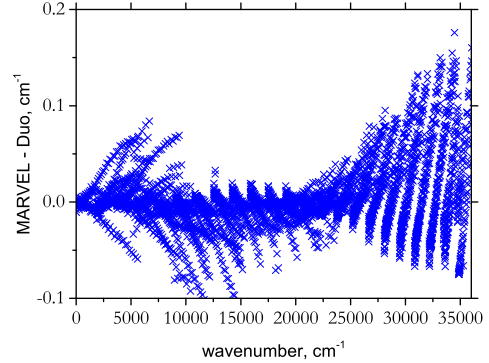}
\caption{Obs. - Calc. residuals of the final energy level fit, where the calculated values are the refined energies calculated by \textsc{Duo} and the observed values are the \textsc{MARVEL} energies and experimental frequencies.}
\label{fig:O-C}
\end{figure}

\begin{table}
\caption{Parameters for the refined potential energy and spin-orbit curves, modelled using the Extended Morse Oscillator function, see Eq.~(\ref{eq:EMO}).}
\label{table:P_S_params}
 \begin{tabular}{lll}
 \hline
  Parameter 	&	Potential Energy Curve	&	Spin Orbit Curve	\\
  \hline\hline
$V_{\rm e}$  (cm$^{-1}$)	&	0	&	61.79345640612	\\
$D_{\rm e}$  (cm$^{-1}$)	&	52495.307750971	&	26.97757067068	\\
$r_{\rm e}$  (\AA)	&	1.1507863151853	&	1.2	\\
${p}$	&	4	&	4	\\
${N_l}^a$	&	2	&	1	\\
${N_r}^a$	&	8	&	4	\\
${B_0}$	&	2.76573276212320	&	1.3782871519399	\\
${B_1}$	&	0.17739962868001	&	0.2416353146388	\\
${B_2}$	&	0.12996658564591	&	-2.4682888462767	\\
${B_3}$	&	1.81747768030430	&	5.5161066471770	\\
${B_4}$	&	-9.76786082439320	&		\\
${B_5}$	&	32.55261795679300	&		\\
${B_6}$	&	-57.64002246220800	&		\\
${B_7}$	&	55.24637383442700	&		\\
${B_8}$	&	-21.23174396925500	&		\\
  \hline
\end{tabular}

$^a$ The upper bound parameter $N$ in Eq.~(\protect\ref{eq:beta}) is defined as
$N=N_l$ for $r\le r_{\rm e}$ and $N=N_r$ for  $r>r_{\rm e}$.

\end{table}

\begin{table}
\caption{Parameters for the refined spin-rotation and $\Lambda$-doubling expressions, modelled using the \u{S}urkus polynomial expansion, see Eq.~(\ref{eq:bobleroy}).}
\label{table:sr_L_params}
 \begin{tabular}{lll}
 \hline
Parameter 	&	Spin-Rotation	&	$\Lambda$-$[p+2q]$	\\
\hline\hline
$r_{\rm e} (\AA)$	&	1.1507863151853	&	1.1507863151853	\\
${p}$	&	4	&	3	\\
${N}$	&	1	&	1	\\
${A_0}$	&	-0.0047317777443454	&	0.0061357379499906	\\
${A_1}$	&	-0.0175102918408170	&	-0.0057848693802985	\\
\hline
\end{tabular}
\end{table}


Experimental transition frequencies were introduced to the data set
at the final stage of the fitting procedure in order to generate the most accurate set of parameters possible.
The final residuals from fitting the \textsc{Duo} energy levels are plotted in Fig.~\ref{fig:O-C}. Notably, the
largest residuals originate from energy levels with high $J$ and $\varv$, in particular $J>35.5$ and $\varv>14$.
Although some residuals have values ranging up to 0.13~\cm,
80.0\% of the fitted frequencies have Obs.$-$Calc. values of $\leq 0.02$~\cm,
yielding a root-mean-square (rms) of 0.015~\cm.
Complete \Duo\ input and output files are provided in the supplementary information which also include the fitted PEC and SOC.

\subsection{\textsc{SPFIT}: Determination of NO spectroscopic parameters}
\label{section_NO_spec-parameters}

Rotational and rovibrational transitions for NO and its isotopologues
were fitted simultaneously in order to determine an accurate set of spectroscopic parameters
for the X~$^2\Pi$ electronic ground state of NO. In an earlier study by \citet{15MuKoTa.NO}, only
rotational and heterodyne infrared measurements of the main isotopic
species were taken into account. Here
Dunham-type parameters along with some parameters describing
the breakdown of the Born-Oppenheimer approximation \citep{73Watson,80Watson}
were determined for all isotopologues in one fit using the diatomic
Hamiltonian outlined by \citet{radi_Hamiltonian_1979} which also provides
isotopic dependences for the lowest order parameters required to fit a
$^2\Pi$ diatomic radical. More details were given for example in a study
of the BrO radical \citep{BrO_rot_2001}.

Fitting and prediction of the spectra, here as well as previously,
were carried out using programs \textsc{SPFIT} and \textsc{SPCAT}
\citep{spfit_1991}. These programs employ Hund's case (b) quantum
numbers throughout which is appropriate at higher rotational quantum
numbers.  Conversion of Hund's case (b) quanta to case (a) or vice
versa depends on the magnitude of the rotational energy relative to
the magnitude of the spin-orbit splitting.  For $2B(J - 0.5)(J + 0.5)
< |A|$, levels with $J + 0.5 = N$ correlate with $^2 \Pi _{1/2}$ and
levels with $J - 0.5 = N$ correlate with $^2 \Pi _{3/2}$; for larger
values of $J$, the correlation is reversed. The reversal occurs
between $J = 5.5$ and $J = 6.5$ in the case of the NO isotopologues.
Atomic masses were taken from the 2012 Atomic Mass Evaluation (AME)
\citep{AME_2012} which takes into account fairly recent mass
determinations for $^{14}$N \citep{Mass-14N_2004}, $^{18}$O
\citep{Mass-18O_2009} and $^{17}$O \citep{Mass-17O_2010}.

A large body of ground state rotational data involving almost all stable NO isotopologues
was used in the previous work. The isotopic species are NO
\citep{70Neumann.NO,72MeDyxx.NO,76Meerts.NO,77DaJoMc.NO,81LoMcVe.NO,79PiCoWa.NO,99VaStEv.NO},
$^{15}$NO \citep{72MeDyxx.NO,91SaYaWi.NO,99VaStEv.NO},
N$^{18}$O \citep{91SaYaWi.NO,15MuKoTa.NO} and N$^{17}$O and
$^{15}$N$^{18}$O \citep{94SaLiDo.NO}.
Included were also $\varv = 1$ $\Lambda$-doubling transitions
\citep{77DaJoMc.NO,81LoMcVe.NO} and $\varv = 1 - 0$ heterodyne infrared
data \citep{86HiWeMa.NO,96SaMeWa.NO} for the main isotopic species.
The experimental uncertainties were critically evaluated; the reported uncertainties were
employed in the fits in most cases.
Frequency errors in spectroscopic measurements, possibly a consequence of
misassignments, are not uncommon. \citet{08AsRiMu} and
\citet{10Amano} revealed frequency errors of $\sim$60~MHz in the
$J = 1 - 0$ transitions of H$_2$D$^+$ and CH$^+$, respectively, from earlier
measurements. These were obtained by new laboratory measurements. Frequency
errors of $\sim$15~MHz were found in two hyperfine structure lines of SH$^+$
by radio astronomical observations and spectroscopic fitting
\citep{14MuGoCe} and confirmed by more recent laboratory
measurements \citep{15HaZixx}. More subtle, because almost within
the estimated uncertainties, were frequency errors in a large line list
of the SO radical which were caused by the failure of a frequency standard
\citep{96KlSaBe.SO}. In the present case, uncertainties were reduced (for a small number of lines)
or increased (for a very small number of lines) if the reported uncertainties of a (sub-)
set of transition frequencies appeared to be judged too conservatively or too optimistically,
respectively. If the reported uncertainties were deemed to be appropriate for the most part,
but few lines had rather large residuals in the fits (usually more than four times the
uncertainties), these lines were omitted.

Extensive sets of Fourier transform infrared data from several sources were
added to the line list in the present study. We used commonly the most
accurate data in cases of multiple studies with essentially the same
quantum number coverage.

The initial spectroscopic parameters \citep{15MuKoTa.NO} reproduced the NO
$\varv = 1 - 0$ data of \citet{94SpChGi.NO} well, and the spectroscopic parameters
barely changed after the fit. We used the reported uncertainties in the fit and included
the $\Lambda$-doubling as far as it had been resolved experimentally. Four lines, $Q(0.5)f$
and $e$ of $^2 \Pi _{1/2}$ and $R(11.5)$ and $R(19.5)$ of $^2 \Pi _{3/2}$, however, showed
residuals between five and almost ten times the reported uncertainties. The remaining lines
were reproduced within the experimental uncertainties both before and after adjustment of
the spectroscopic parameters after these four lines were omitted from the fit. The uncertainties
of some parameters were improved.

The NO $\varv = 1 - 0$ data of \citet{95CoDaMa.NO} had some overlap with the rovibrational
data already in the fit. A trial fit suggested that these data were 0.00012~cm$^{-1}$ too high,
a considerable fraction of the reported uncertainties ranging from 0.00015 to 0.00022~cm$^{-1}$,
with rather small scatter. All transition frequencies were reproduced very well after modifying
the transition frequencies by 0.00012~cm$^{-1}$. The final rms error for these data was only
slightly worse than 0.4, indicating a slightly conservative judgement of the adjusted data.
This data set was the only one for which the line positions were adjusted. In all other data
sets pertaining to the main isotopic species we did not find any clear evidence for possible
calibration errors.

The $\varv = 2 - 1$ data of \citet{97MaDaRe.NO} were added next with their reported
uncertainties.
In no other instances were uncertainties specified; these were estimated to be reproduced within uncertainties, on average, in the final fits.
In the case of
a small number of lines in a data set with residuals much larger than most other lines,
these lines were omitted, and the uncertainties were evaluated on the basis of the remaining
lines. Uncertainties of 0.0002 and 0.0003~cm$^{-1}$ were assumed for the $\varv = 2 - 0$
data of \citet{94DaMaCo.NO} in the $^2 \Pi _{1/2}$ and $^2 \Pi _{3/2}$ spin ladder,
respectively. We assumed an uncertainty of 0.0005~cm$^{-1}$ for the $\varv = 1 - 0$ and $2 - 1$ data
of \citet{NO_v1-0_2-1_plus_1978}.

The extensive $\Delta \varv = 2$ data of \citet{80AmVexx.NO} were
included next, followed by the $\varv = 3 - 0$ data of \citet{78HeLeCa.NO} and the
$\Delta \varv = 3$ data of \citet{82Amiot.NO}. An additional constraint
was a smooth trend from lower $\varv$ to higher $\varv$ for two extensive sets of data.
Uncertainties were between 0.0005~cm$^{-1}$ for $\varv = 2 - 0$ and a few more to
0.0024~cm$^{-1}$ for $\varv = 15 - 13$ of \citet{80AmVexx.NO}.
We applied 0.0007~cm$^{-1}$ for the data of \citet{78HeLeCa.NO} and from
0.002~cm$^{-1}$ for $\varv = 10 - 7$ and a few more to 0.004~cm$^{-1}$ for
$\varv = 22 - 19$ of \citet{82Amiot.NO}.

The quality of the $\Delta \varv = 2$ data of \citet{79HaJoLe.NO} up to
$\varv = 6 - 4$ was questioned by \citet{80AmVexx.NO} because of the
low resolution and the large deviations of the transition frequencies. The $P$-branch
transition assignments in $\varv = 2 - 0$ are essentially complete up to $J = 65.5$
with additional assignments reaching $J = 77.5$. Transition frequencies up to $J = 64.5$
could be reproduced to 0.008~cm$^{-1}$, but the impact of these data on the spectroscopic
parameter values and uncertainties was negligible. Higher-$J$ data were too sparse and
showed very large residuals with some scatter that could not be reduced sufficiently
with parameter values that were deemed reasonable. The higher-$\varv$ data from that
work showed even larger scatter in the residuals, such that the data of
\citet{79HaJoLe.NO} were omitted entirely.

Overtone spectra involving larger $\Delta \varv$ involved transition frequencies too
limited in $J$ and in accuracy such that we did not consider these data.

Data for $^{15}$NO and $^{15}$N$^{18}$O were taken from \citet{80TeHeCa.NO};
the uncertainties used for $^{15}$NO were 0.0005, 0.0010 and 0.0015~cm$^{-1}$
for $\varv = 1 - 0$, $2 - 0$ and $3 - 0$, respectively, and slightly lower for the two
overtone bands of $^{15}$N$^{18}$O. Additional data for $^{15}$NO, $\varv = 1 - 0$
and $2 - 1$, as well as the $\varv = 1 - 0$ bands of  $^{15}$N$^{17}$O and
$^{15}$N$^{18}$O were taken from \citet{565758_IR_1-0_+_1979} with uncertainties
of 0.0003~cm$^{-1}$.


\begin{table}
  \begin{center}
  \caption{Present and previous spectroscopic parameters$^a$ (MHz) for NO determined
           from the isotopic invariant fit.}
  \label{table_NO_spec-parameters}
  \begin{tabular}{lr@{}lr@{}l}
  \hline \hline
Parameter                                                             & \multicolumn{2}{c}{Present}  & \multicolumn{2}{c}{Previous} \\[1pt]
\hline
$U_{10}\mu^{-1/2} \times 10^{-3}$$^b$                                 &     57\,081&.238\,9~(52)   &     56\,240&.216\,66~(14) \\
$U_{10}\mu^{-1/2}{\it
{\Delta}} _{10}^{\rm N}m_e/M_{\rm N}$                &      2\,104&.2~(54)        &            &              \\
$U_{10}\mu^{-1/2}{\it
{\Delta}} _{10}^{\rm O}m_e/M_{\rm O}$                &         573&.6~(56)        &            &              \\
$Y_{20} \times 10^{-3}$                                               &      $-$422&.325\,96~(105) &            &              \\
$Y_{30}$                                                              &         293&.06~(33)       &            &              \\
$Y_{40}$                                                              &        $-$3&.551~(49)      &            &              \\
$Y_{50}$                                                              &        $-$0&.364\,6~(37)   &            &              \\
$Y_{60} \times 10^3$                                                  &        $-$3&.264~(134)     &            &              \\
$Y_{70} \times 10^3$                                                  &        $-$0&.191\,75~(192) &            &              \\
$U_{01}\mu^{-1}$                                                      &     51\,119&.462\,5~(41)   &     51\,119&.680\,7~(42)  \\
$U_{01}\mu^{-1}{\it
{\Delta}} _{01}^{\rm N}m_e/M_{\rm N}$                  &        $-$4&.530\,8~(29)   &        $-$4&.469\,2~(29)  \\
$U_{01}\mu^{-1}{\it
{\Delta}} _{01}^{\rm O}m_e/M_{\rm O}$                  &        $-$4&.082\,0~(27)   &        $-$4&.027\,2~(27)  \\
$Y_{11}$                                                              &      $-$525&.876\,0~(20)   &      $-$526&.763\,3~(22)  \\
$Y_{21}$                                                              &        $-$0&.433\,78~(145) &            &              \\
$Y_{31} \times 10^3$                                                  &        $-$4&.92~(33)       &            &              \\
$Y_{41} \times 10^3$                                                  &        $-$0&.817~(35)      &            &              \\
$Y_{51} \times 10^6$                                                  &           1&.34~(170)      &            &              \\
$Y_{61} \times 10^6$                                                  &        $-$1&.323~(30)      &            &              \\
$U_{02}\mu^{-2} \times 10^3$                                          &      $-$163&.955\,7~(23)   &      $-$163&.944\,1~(30)  \\
$U_{02}\mu^{-2}{\it
{\Delta}} _{02}^{\rm N}m_e/M_{\rm N} \times 10^3$      &           0&.044\,0~(23)   &           0&.044\,7~(24)  \\
$Y_{12} \times 10^3$                                                  &        $-$0&.451\,88~(96)  &        $-$0&.484\,2~(55)  \\
$Y_{22} \times 10^6$                                                  &       $-$15&.24~(47)       &            &              \\
$Y_{32} \times 10^6$                                                  &           0&.696~(58)      &            &              \\
$Y_{42} \times 10^6$                                                  &        $-$0&.100\,7~(20)   &            &              \\
$Y_{03} \times 10^9$                                                  &          41&.282~(182)     &          37&.940~(114)    \\
$Y_{13} \times 10^9$                                                  &        $-$6&.74~(28)       &            &              \\
$A_{00}^{\rm BO} \times 10^{-3}$                                      &      3\,695&.038\,00~(69)  &      3\,695&.104\,22~(65) \\
$A_{00}^{\rm BO}{\it
{\Delta}} _{00}^{A{\rm ,N}}m_e/M_{\rm N}$             &         186&.24~(27)       &         204&.98~(26)      \\
$A_{00}^{\rm BO}{\it
{\Delta}} _{00}^{A{\rm ,O}}m_e/M_{\rm O}$             &         151&.30~(38)       &         167&.83~(38)      \\
$A_{10}$                                                              &   $-$7\,069&.06~(95)       &   $-$7\,335&.247~(55)     \\
$A_{20}$                                                              &      $-$123&.86~(62)       &            &              \\
$A_{30}$                                                              &        $-$5&.757~(105)     &            &              \\
$A_{40} \times 10^3$                                                  &          52&.0~(68)        &            &              \\
$A_{50} \times 10^3$                                                  &       $-$11&.084~(147)     &            &              \\
$A_{01}$                                                              &           0&.124\,8~(59)   &           0&.122\,8~(59)  \\
$\gamma_{00}$                                                         &      $-$193&.05~(21)       &      $-$193&.40~(21)      \\
$\gamma_{10}$                                                         &           6&.741~(46)      &           7&.476\,3~(55)  \\
$\gamma_{20}$                                                         &           0&.345~(30)      &            &              \\
$\gamma_{30} \times 10^3$                                             &        $-$7&.3~(33)        &            &              \\
$\gamma_{40} \times 10^3$                                             &           1&.512~(105)     &            &              \\
$\gamma_{01} \times 10^3$                                             &           1&.530\,0~(133)  &           1&.611\,0~(56)  \\
$\gamma_{11} \times 10^3$                                             &           0&.164~(24)      &            &              \\
$p_{00}^{\rm BO,eff}$                                                 &         350&.623\,39~(91)  &         350&.623\,40~(91) \\
$p_{00}^{\rm BO}{\it
{\Delta}} _{00}^{p{\rm ,N}}m_e/M_{\rm N} \times 10^3$ &       $-$17&.12~(93)       &       $-$17&.11~(93)      \\
$p_{10} \times 10^3$                                                  &      $-$403&.50~(32)       &      $-$403&.50~(32)      \\
$p_{01} \times 10^6$                                                  &          34&.1~(12)        &          34&.1~(12)       \\
$q_{00}$                                                              &           2&.844\,718~(39) &           2&.844\,711~(39)\\
$q_{10} \times 10^3$                                                  &       $-$44&.283~(65)      &       $-$44&.282~(65)     \\
$q_{01} \times 10^6$                                                  &          42&.313~(112)     &          42&.319~(112)    \\
\hline \hline
\end{tabular}\\[2pt]
\end{center}
{
$^a$ Numbers in parentheses are 1\,$\sigma$ uncertainties in units of the least
     significant figures. Previous parameter values from \citet{15MuKoTa.NO}.
$^b$ Previous value corresponds to an effective $Y_{10} \times 10^{-3}$.
}
\end{table}


\begin{table}
  \begin{center}
  \caption{Present and previous hyperfine parameters$^a$ (MHz) for NO determined
           from the isotopic invariant fit.}
  \label{table_NO_HFS-parameters}
  \begin{tabular}{lr@{}lr@{}l}
  \hline \hline
Parameter                                                             & \multicolumn{2}{c}{Present}  & \multicolumn{2}{c}{Previous} \\[1pt]
\hline
$a_{00}({\rm N})$                                                     &          84&.304\,2~(106)  &          84&.304\,2~(106) \\
$a_{10}({\rm N}) \times 10^3$                                         &      $-$202&.3~(211)       &      $-$202&.3~(211)      \\
$b_{F,00}({\rm N})$                                                   &          22&.270~(21)      &          22&.271~(21)     \\
$b_{F,10}({\rm N}) \times 10^3$                                       &         250&.~(43)         &         249&.~(43)        \\
$c_{00}({\rm N})$                                                     &       $-$58&.890\,4~(14)   &       $-$58&.890\,4~(14)  \\
$d_{00}({\rm N})$                                                     &         112&.619\,48~(132) &         112&.619\,47~(132)\\
$d_{10}({\rm N}) \times 10^3$                                         &       $-$30&.3~(27)        &       $-$30&.3~(27)       \\
$d_{01}({\rm N}) \times 10^6$                                         &         105&.6~(145)       &         105&.6~(145)      \\
$eQq_{1,00}({\rm N})$                                                 &        $-$1&.898\,6~(32)   &        $-$1&.898\,6~(32)  \\
$eQq_{1,10}({\rm N}) \times 10^3$                                     &          77&.4~(64)        &          77&.4~(64)       \\
$eQq_{2,00}({\rm N})$                                                 &          23&.112\,6~(62)   &          23&.112\,6~(62)  \\
$eQq_{S,00}({\rm N}) \times 10^3$                                     &        $-$6&.89~(83)       &        $-$6&.89~(83)      \\
$C_{I,00}({\rm N}) \times 10^3$                                       &          12&.293~(27)      &          12&.293~(27)     \\
$C'_{I,00}({\rm N}) \times 10^3$                                      &           7&.141~(123)     &           7&.141~(123)    \\
$a_{00}({\rm O})$                                                     &      $-$173&.058\,3~(101)  &      $-$173&.058\,3~(101) \\
$b_{F,00}({\rm O})$                                                   &       $-$35&.458~(109)     &       $-$35&.460~(109)    \\
$c_{00}({\rm O})$                                                     &          92&.868~(171)     &          92&.871~(171)    \\
$d_{00}({\rm O})$                                                     &      $-$206&.121\,6~(70)   &      $-$206&.121\,6~(70)  \\
$eQq_{1,00}({\rm O})$                                                 &        $-$1&.331~(47)      &        $-$1&.330~(47)$^b$ \\
$eQq_{2,00}({\rm O})$                                                 &       $-$30&.01~(163)      &       $-$30&.02~(163)     \\
$C_{I,00}({\rm O}) \times 10^3$                                       &       $-$32&.7~(23)        &       $-$32&.7~(23)       \\
\hline \hline
\end{tabular}\\[2pt]
\end{center}
{
$^a$ Numbers in parentheses are 1\,$\sigma$ uncertainties in units of the least
     significant figures. Previous parameter values from \citet{15MuKoTa.NO}.
$^b$ Small error in value corrected.
}
\end{table}


\begin{table}
  \begin{center}
  \caption{Derived parameters (MHz, pm, unitless)$^a$ of NO from the isotopic invariant fit.}
  \label{table_derived-parameters}
\smallskip
  \begin{tabular}{lr@{}lr@{}l}
  \hline \hline
Parameter                           & \multicolumn{2}{c}{Present}    & \multicolumn{2}{c}{Previous}  \\[1pt]
\hline
$Y_{10} \times 10^{-3}$             &     57\,083&.936\,69~(89)    &            &               \\
${\it
{\Delta}} _{10}^{\rm N}$           &           0&.941\,0~(24)     &            &               \\
${\it
{\Delta}} _{10}^{\rm O}$           &           0&.303\,2~(29)     &            &               \\
$Y_{01}$                            &     51\,110&.849\,70~(68)    &     51\,111&.184\,2~(11)   \\
${\it
{\Delta}} _{01}^{\rm N}$           &        $-$2&.262\,44~(146)   &        $-$2&.231\,66~(147) \\
${\it
{\Delta}} _{01}^{\rm O}$           &        $-$2&.328\,23~(156)   &        $-$2&.296\,99~(156) \\
$B_e$                               &     51\,110&.888\,44~(76)    &            &               \\
$r_e$                               &         115&.078\,792\,9~(9) &            &               \\
$Y_{02} \times 10^3$                &      $-$163&.911\,79~(49)    &      $-$163&.899\,4~(27)   \\
${\it
{\Delta}} _{02}^{\rm N}$           &        $-$6&.84~(35)         &        $-$6&.96~(37)       \\
$A_{00}$                            & 3\,695\,375&.54~(39)         & 3\,695\,477&.03~(21)       \\
${\it
{\Delta}} _{00}^{A{\rm ,N}}$       &           1&.286\,6~(18)     &           1&.416\,0~(18)   \\
${\it
{\Delta}} _{00}^{A{\rm ,O}}$       &           1&.193\,9~(30)     &           1&.324\,3~(30)   \\
$p_{00}$                            &         350&.606\,27~(17)    &         350&.606\,29~(17)  \\
${\it
{\Delta}} _{00}^{p{\rm ,N}}$       &        $-$1&.246~(68)        &        $-$1&.246~(68)      \\
\hline \hline
\end{tabular}\\[2pt]
\end{center}
{
$^a$ Numbers in parentheses are 1\,$\sigma$ uncertainties in units of the least significant figures.
     $r_e$ in pm, ${\it
{\Delta}}$s unitless, all other parameters in MHz. Previous parameter values from
     \citet{15MuKoTa.NO}; empty fields indicate values have or could not be determined
     except for the previous effective $Y_{10}$ value which was devoid of all vibrational corrections.
}
\end{table}


Our Hamiltonian for NO has been described earlier \citep{15MuKoTa.NO}; however,
in order to fit the FTIR data pertaining to the main isotopic species, we had to add
several vibrational corrections to the mechanical and fine-structure parameters.
These parameters were carefully chosen at each step of the fitting procedure by
searching among the reasonable parameters for the one that reduces the rms error of
the fit the most. A new parameter led sometimes to large changes in the value of one or more
spectroscopic parameters. Such a parameter was kept in the fit only if additional transition
frequencies did not lead to drastic changes in the value of this parameter. If two
parameters led to similar reductions in the rms error and both together led to a much larger
reduction than either one, both parameters were kept in the fit; the decision was
postponed otherwise.

The isotopic FTIR data required Born-Oppenheimer breakdown parameters
to the lowest order vibrational parameter ($Y_{10}$) to be added.
Other Born-Oppenheimer breakdown parameters were barely determined, at best, and were
omitted from the final fits.
The final spectroscopic parameters are
given in Tables~\ref{table_NO_spec-parameters} and \ref{table_NO_HFS-parameters}, derived parameters are in Table~\ref{table_derived-parameters}, in both cases presented alongside data from the previous study \citep{15MuKoTa.NO}.
The reported
uncertainties are only those from the respective fits; uncertainties
of the atomic masses \citep{AME_2012} (for the ${\it {\Delta}}$s and
for $r_e$), the mass of the electron in atomic mass units (for the
${\it {\Delta}}$s), or of the conversion factor from $B_e$ to the
moment of inertia, derived from \citet{fundamental_constants_2012}
(see also \citet{SiO_rot_2013} for the conversion factor), are
negligible here. The line, parameter and fit files will be available
in the
CDMS\footnote{http://www.astro.uni-koeln.de/site/vorhersagen/pickett/beispiele/NO/}
\citep{cdms}. The comparison between present and previous
spectroscopic parameters is frequently quite favourable. The addition
of new parameters due to new data can lead to changes outside the
combined uncertainties; such changes can even be relatively large in
cases in which a lower order parameter is comparatively small in
magnitude with respect to the magnitude of a higher order parameter.
An example for the latter case is $A_{10}$, examples for the former
are the related changes in $A_{00}^{\rm BO}$ and its Born-Oppenheimer
breakdown parameters.

Sensitive overtone measurements of NO isotopologues, similar to those
carried out for CO $\varv = 3 - 0$ \citep{CO-isos_3-0_2015} and $\varv
= 4 - 0$ \citep{CO-isos_4-0_2015}, are probably the most
straightforward way to improve the NO spectroscopic parameters and
predictions of rovibrational spectra, especially those of minor
isotopic species.

Predictions based on the present set of spectroscopic parameters (generated with {\sc SPCAT})  should be quite good up to $J$ of around 60 or 70 for low values of $\varv$ and for the main isotopic
species, but considerable caution is advised beyond $J$ of 90. The quality of the
predictions is expected to deteriorate somewhat toward $\varv = 20$. The vibrational
states $\varv = 20$, 21 and 22 are at the edge of the data set; predictions involving
these states should be reasonable. Extrapolation in $\varv$ should be viewed with more
 caution; data involving $\varv = 25$ may be reasonable. By comparing to the corresponding \Duo\ values, which were obtained from an independent fit, the prediction error of these two methods should be within 0.07~\cm\ for  $\varv = 25$ and not exceed 1~\cm\ for $\varv = 27$. Using the same argument for rotational excitations, we find that the difference between the $J=99.5$ energies obtained with two methods grow from 0.1~\cm\ ($v=0$, $\tilde{E}=16,300$~\cm) to 9.5~\cm\ ($v=20$, $\tilde{E}=41,300$~\cm) and then to 96~\cm\ ($v=27$, $\tilde{E}=51,500$~\cm). These difference give an indication both of the \textsc{SPCAT} and \Duo\ extrapolation errors at high $v$ and $J$.

On the basis of the available data, we expect predictions for $^{15}$NO to be
slightly less reliable, and those of isotopologues involving $^{18}$O or $^{17}$O somewhat
less reliable still at low values of $\varv$. Moreover, predictions involving $\varv = 5$
and higher should be viewed with considerable caution.

\subsection{Duo: Line List}	\label{Duo_linelist}

\subsubsection{Line List Calculations}	\label{linelist}

The line list computed in \textsc{Duo} comprises of two files
\citep{jt631}; the \texttt{.states} file contains the running number,
line position (\cm) (i.e. energy term value), total statistical weight
and associated quantum numbers.
The \texttt{.states} file also includes lifetimes for each state and Land\'e $g$-factors.
The \texttt{.trans} file
contains the upper and lower level running number, Einstein-A
coefficients (s$^{-1}$) and transition wavenumber (\cm).
The Einstein-A coefficient is the rate of
spontaneous emission between the upper and lower energy levels.


The NO ground electronic state line list was computed with
\textsc{Duo} using the nuclear statistical weights $g_{\rm ns} =
(2I_{\rm N} + 1)(2I_{\rm O} + 1)$, where $I_{\rm I}$ and $I_{\rm O}$
are the nuclear spins of the nitrogen (1 for $^{14}$N and $1/2$ for
$^{15}$N) and oxygen (0 for $^{16}$O and $^{18}$O and $5/2$ for
$^{17}$O) atoms, respectively.

The
complete $^{14}$N$^{16}$O line list contains 21,688
states and   2,281,042 transitions in the wavenumber
range 0 - 40,000 \cm, extending to a maximum rotational quantum number
of 184.5 and a maximum vibrational quantum number of 51;
an extract of the \texttt{.states} and \texttt{.trans} files
are shown in Tables \ref{table:states} and \ref{table:trans},
respectively.

Line lists for the six combinations of $^{14}$N, $^{15}$N, $^{16}$O,
$^{17}$O and $^{18}$O were computed, without any adjustments to the
fit; only the masses were altered to the values specified above.

\begin{figure}
\centering
  \centering
  \includegraphics[scale=0.3]{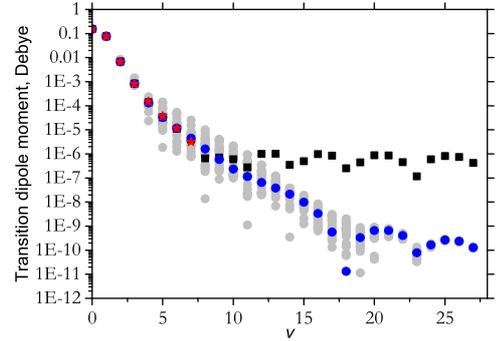}
\caption{Vibrational transition dipole moments (D) from the $\varv=0$ ground state.  of $^{14}$N$^{16}$O: empirical \citep{06LeChOg.NO} (stars) and \ai\ calculated using the quintic splines (squares) and Pad\'{e}-type expansion (circles).}
  \label{fig:TDM}
\end{figure}

In order to avoid the numerical noise associated with the small dipole moment matrix
elements \citep{15LiGoRo.CO}, we followed the suggestion of
\citet{16MeMeSt} and represented the \ai\ dipole moment using an
analytical function. To this end the following Pad\'{e} form due to
\citet{63Goodisman} was used:
\begin{equation}\label{e:Pade}
\mu(r) = \frac{z^3}{1+z^7}\sum_{i \ge 0} a_i T_i\left(\frac{z-1}{z+1}\right),
\end{equation}
where $z=r/r_{\rm e}$, $T_i(x)$ are Chebyshev polynomials and $a_i$
are expansion parameters obtained by fitting to 352 \ai\ dipole moment
values covering $r = 0.7 - 9 $~\AA. With 18 parameters we were able to
reproduce the \ai\ dipole with a root-mean-square (rms) error of
0.07~D for the whole range, with best agreement in the vicinity of the
equilibrium of the order $10^{-5} - 10^{-6}$~D.  The vibrational
transition moments computed using the quintic splines interpolation
implemented as default in \Duo\ and this Pad\'{e} expression are shown in
Fig.~\ref{fig:TDM}, where they are also compared to the empirical
values, see \citet{06LeChOg.NO} and references therein, where
available. They are also listed in Table~\ref{t:TDM}. The
spline-interpolated dipoles produce an artificial plateau-like error
of $10^{-6}$--$10^{-7}$~D as expected \citep{15LiGoRo.CO}. The
analytical form improves this by shifting the error to
$10^{-10}$--$10^{-11}$~D. However, the transition dipole moment values
appear to be very sensitive to such functional interpolation, at least
within $10^{-5} - 10^{-6}$~D, which is also the absolute error of our
interpolation scheme. To illustrate this, Fig.~\ref{fig:TDM} also
shows vibrational transition dipole moments, computed using fits
with different expansion orders, ranges, weighings of the data, etc.
From all these combinations we then selected the
set which gives the closest agreement with the transition dipole moments obtained using the
spline-interpolation scheme. This set is also in the best agreement with the
empirical transition dipole moments.

In intensity (line list) calculations we used a dipole threshold of
$1\times10^{-9}$~D, i.e. all vibrational matrix elements smaller than
this value were set to zero to avoid artificial intensities due to the
numerical error.

\begin{table*}
\caption{ Transition dipole moments of NO. The total uncertainties are given in parenthesis. The \ai\  values are obtained using the \ai\ DMC interpolated with the quintic splines and Pad\'{e} expression as in Eq.~(\ref{e:Pade}).}
\label{t:TDM}
 \begin{tabular}{rccllrr}
 \hline
  Band   &          &        & `Exp'           &  Ref.                   &  Splines&  Pad\'{e}  \\
 \hline
  0--   0&          &        &    0.1595(15)   & \citet{82Amiot.NO} &     0.155&     0.155  \\ %
  1--   0& $ 10^{-2}$&$\times$ &    7.6931(14)   & \citet{95CoDaMa.NO} &    7.649&     7.646  \\ %
  2--   0& $ 10^{-3}$&$\times$ &      6.78(20)   & \citet{97MaDaRe.NO} &    6.865&     6.890  \\ %
  3--   0& $ 10^{-4}$&$\times$ &     7.975(23)   & \citet{06LeChOg.NO} &    8.372&     8.379  \\ %
  4--   0& $ 10^{-4}$&$\times$ &    1.4804(45)   & \citet{06LeChOg.NO} &    1.396&     1.300  \\ %
  5--   0& $ 10^{-5}$&$\times$ &     3.683(17)   & \citet{06LeChOg.NO} &    3.319&     3.244  \\ %
  6--   0& $ 10^{-5}$&$\times$ &     1.136(06)   & \citet{06LeChOg.NO} &    1.100&     1.182  \\ %
  7--   0& $ 10^{-6}$&$\times$ &      3.09(47)   & \citet{06BoMcOs.NO} &    3.959&     4.458  \\ %
  3--   1& $ 10^{-2}$&$\times$ &      1.19(12)   & \citet{98MaDaRe.NO} &         &     1.194  \\ %
  2--   1&          &        &     0.109(38)   & \citet{94DaMaCo.NO} &         &     0.108  \\ %
  7--   6& $ 10^{-1}$&$\times$ &      1.89(11)   & \citet{97DrWoxx.NO} &         &     1.965  \\ %
 21--  20& $ 10^{-1}$&$\times$ &     3.176(82)   & \citet{97DrWoxx.NO} &         &     3.015  \\ %
 21--  19& $ 10^{-1}$&$\times$ &     1.077(27)   & \citet{97DrWoxx.NO} &         &     1.047  \\ %
 21--  18& $ 10^{-2}$&$\times$ &      3.68(16)   & \citet{97DrWoxx.NO} &         &     3.220  \\ %
 21--  17& $ 10^{-2}$&$\times$ &      1.09(16)   & \citet{97DrWoxx.NO} &         &     1.239  \\ %
\hline
\end{tabular}
\end{table*}

Our final \Duo\ input, which defines our final PEC, SOC and DMC as
well as other input parameters selected, is given in the supplementary data.

\subsubsection{Hybrid Line List}	\label{hybrid}

The final lists were produced by combining the {\sc SPCAT} frequencies and
\Duo\ Einstein coefficients. To this end we used the advantage of the
two-file structure of the ExoMol format \citep{jt631}, with \texttt{.states}  and
\texttt{.trans} files. We simply replaced the \Duo\ energies in the states
file with the corresponding {\sc SPCAT} values. The corresponding coverage of
{\sc SPCAT} and \Duo\ is summarized in Table~\ref{table:iso}.  The correlation based on the
Hund's case~(a) quantum numbers ($J, \varv, \Omega$ and parity) was
straightforward. The \Duo\ energies extend significantly beyond the  {\sc SPCAT} data range.
In order to prevent possible jumps and discontinuities when switching between these data set,
the \Duo\ energies were shifted to match the {\sc SPCAT} energies at the points of the switch.
For example, in case of \NO{14}{16}, the maximum vibrational excitation considered by {\sc SPCAT} $\varv_{\rm max}$ is 29 (the switching point),
therefore all  \Duo\ energies for $v=29\ldots 51$ where shifted such that \Duo\ $\varv = 29$ energy value coincides with those by
{\sc SPCAT} for all $J$s.
The same strategy was used to stitch the {\sc SPCAT} and \Duo\ energies at $J =  99.5$ (the chosen threshold for \textsc{SPCAT}):
the \Duo\ values for $J = 99.5 \ldots 185.5$ were shifted to match the corresponding $J = 99.5$ value of {\sc SPCAT} for each $\varv$, $\Omega$ and parity individually.

The {\sc SPCAT} energies of the isotopologues are  even more limited in terms of the vibrational coverage;
$\varv_{\rm max}$ = 9 (\NO{15}{16} and \NO{14}{18}) and $\varv_{\rm max}$ = 4
(\NO{14}{17}, \NO{15}{17} and \NO{15}{18}).  As for the main isotopologue, we used the corresponding \Duo\ energies
to top up the corresponding line list to the same thresholds as for \NO{14}{16}. The better representation of the
data from the parent isotopologue helped us to improve the accuracy of the \Duo\ prediction for $v \le 29$.
By comparing the {\sc SPCAT} and \Duo\ energies of \NO{14}{16} in this range, the corresponding residuals were
propagated (for each rovibronic state individually) to correct the corresponding \Duo\ energies for the other five
isotopologues  (see, for example, \citet{jt665}).
The energies for $\varv \ge \varv_{\rm max}$ were then given by:
$$
E_{J,\pm,v,\Omega}^{\rm iso} = E_{J,\pm,v,\Omega}^{\rm Duo-iso} + E_{J,\pm,v,\Omega}^{\rm SPCAT-parent} - E_{J,\pm,v,\Omega}^{\rm Duo-parent},
$$
where `Duo-iso' refers to the \Duo\ energies of one of the five isotopologues for a given set of $J,\pm,v,\Omega$, while
`Duo-parent'  and `\textsc{SPCAT}-parent' indicate the corresponding energies of the parent isotopologue computed by \Duo\ and {\sc SPCAT}, respectively.
The line lists do not include the hyperfine structure of the energy levels and transitions.

\begin{table*}
\caption{Extract from the states file of the $^{14}$N$^{16}$O line list.   }
\label{table:states}
 \begin{tabular}{crccccccrrrrrc}
 \hline
$i$ & Energy (\cm)	&	$g_i$	&	$J$ & $\tau$ & $g_J$	&	Parity	&	e/f	&	State	&	$\varv$	&	${\Lambda}$	&	${\Sigma}$	&	${\Omega}$ & emp/calc	\\
\hline\hline
       1  &        0.000000  &    6  &     0.5  &   inf       &   -0.000767  &     +    &   e    &     X1/2    &      0  &      1  &    -0.5  &     0.5  &      e       \\
       2  &     1876.076228  &    6  &     0.5  &   8.31E-02  &   -0.000767  &     +    &   e    &     X1/2    &      1  &      1  &    -0.5  &     0.5  &      e       \\
       3  &     3724.066346  &    6  &     0.5  &   4.25E-02  &   -0.000767  &     +    &   e    &     X1/2    &      2  &      1  &    -0.5  &     0.5  &      e       \\
       4  &     5544.020643  &    6  &     0.5  &   2.89E-02  &   -0.000767  &     +    &   e    &     X1/2    &      3  &      1  &    -0.5  &     0.5  &      e       \\
       5  &     7335.982597  &    6  &     0.5  &   2.22E-02  &   -0.000767  &     +    &   e    &     X1/2    &      4  &      1  &    -0.5  &     0.5  &      e       \\
       6  &     9099.987046  &    6  &     0.5  &   1.81E-02  &   -0.000767  &     +    &   e    &     X1/2    &      5  &      1  &    -0.5  &     0.5  &      e       \\
       7  &    10836.058173  &    6  &     0.5  &   1.54E-02  &   -0.000767  &     +    &   e    &     X1/2    &      6  &      1  &    -0.5  &     0.5  &      e       \\
       8  &    12544.207270  &    6  &     0.5  &   1.35E-02  &   -0.000767  &     +    &   e    &     X1/2    &      7  &      1  &    -0.5  &     0.5  &      e       \\
       9  &    14224.430238  &    6  &     0.5  &   1.21E-02  &   -0.000767  &     +    &   e    &     X1/2    &      8  &      1  &    -0.5  &     0.5  &      e       \\
      10  &    15876.704811  &    6  &     0.5  &   1.10E-02  &   -0.000767  &     +    &   e    &     X1/2    &      9  &      1  &    -0.5  &     0.5  &      e       \\
      11  &    17500.987446  &    6  &     0.5  &   1.01E-02  &   -0.000767  &     +    &   e    &     X1/2    &     10  &      1  &    -0.5  &     0.5  &      e       \\
      12  &    19097.209871  &    6  &     0.5  &   9.41E-03  &   -0.000767  &     +    &   e    &     X1/2    &     11  &      1  &    -0.5  &     0.5  &      e       \\
      13  &    20665.275246  &    6  &     0.5  &   8.83E-03  &   -0.000767  &     +    &   e    &     X1/2    &     12  &      1  &    -0.5  &     0.5  &      e       \\
      14  &    22205.053904  &    6  &     0.5  &   8.35E-03  &   -0.000767  &     +    &   e    &     X1/2    &     13  &      1  &    -0.5  &     0.5  &      e       \\
      15  &    23716.378643  &    6  &     0.5  &   7.94E-03  &   -0.000767  &     +    &   e    &     X1/2    &     14  &      1  &    -0.5  &     0.5  &      e       \\
      16  &    25199.039545  &    6  &     0.5  &   7.59E-03  &   -0.000767  &     +    &   e    &     X1/2    &     15  &      1  &    -0.5  &     0.5  &      e       \\
      17  &    26652.778266  &    6  &     0.5  &   7.30E-03  &   -0.000767  &     +    &   e    &     X1/2    &     16  &      1  &    -0.5  &     0.5  &      e       \\
      18  &    28077.281796  &    6  &     0.5  &   7.05E-03  &   -0.000767  &     +    &   e    &     X1/2    &     17  &      1  &    -0.5  &     0.5  &      e       \\
      19  &    29472.175632  &    6  &     0.5  &   6.84E-03  &   -0.000767  &     +    &   e    &     X1/2    &     18  &      1  &    -0.5  &     0.5  &      e       \\
      20  &    30837.016339  &    6  &     0.5  &   6.66E-03  &   -0.000767  &     +    &   e    &     X1/2    &     19  &      1  &    -0.5  &     0.5  &      e       \\
      21  &    32171.283479  &    6  &     0.5  &   6.50E-03  &   -0.000767  &     +    &   e    &     X1/2    &     20  &      1  &    -0.5  &     0.5  &      e       \\
      22  &    33474.370850  &    6  &     0.5  &   6.38E-03  &   -0.000767  &     +    &   e    &     X1/2    &     21  &      1  &    -0.5  &     0.5  &      e       \\
      23  &    34745.577033  &    6  &     0.5  &   6.27E-03  &   -0.000767  &     +    &   e    &     X1/2    &     22  &      1  &    -0.5  &     0.5  &      e       \\
      24  &    35984.095189  &    6  &     0.5  &   6.19E-03  &   -0.000767  &     +    &   e    &     X1/2    &     23  &      1  &    -0.5  &     0.5  &      e       \\
      25  &    37189.002091  &    6  &     0.5  &   6.13E-03  &   -0.000767  &     +    &   e    &     X1/2    &     24  &      1  &    -0.5  &     0.5  &      e       \\
      26  &    38359.246347  &    6  &     0.5  &   6.09E-03  &   -0.000767  &     +    &   e    &     X1/2    &     25  &      1  &    -0.5  &     0.5  &      e       \\
      27  &    39493.635791  &    6  &     0.5  &   6.07E-03  &   -0.000767  &     +    &   e    &     X1/2    &     26  &      1  &    -0.5  &     0.5  &      e       \\
\hline

\end{tabular}
\mbox{}\\
{\flushleft
$i$:   State counting number.     \\
$\tilde{E}$: State energy in \cm. \\
$g$:  Total statistical weight, equal to ${g_{\rm ns}(2J + 1)}$.     \\
$J$: Total angular momentum.\\
$\tau$: Lifetime (s$^{-1}$).\\
$g_J$: Land\'{e} $g$-factor \\
$+/-$:   Total parity. \\
$e/f$:   Rotationless parity. \\
State: Electronic state.\\
$\varv$:   State vibrational quantum number. \\
$\Lambda$:  Projection of the electronic angular momentum. \\
$\Sigma$:   Projection of the electronic spin. \\
$\Omega$:   $\Omega=\Lambda+\Sigma$, projection of the total angular momentum.\\
emp/calc:   e= empirical (\textsc{SPCAT}), c=calculated (Duo). \\

}

\end{table*}

\begin{table}
\caption{Extract from the transitions file of the $^{14}$N$^{16}$O  line list. }
\label{table:trans}

\begin{tabular}{rrrr}
\hline
\multicolumn{1}{c}{$f$}	&	\multicolumn{1}{c}{$i$}	& \multicolumn{1}{c}{$A_{fi}$ (s$^{-1}$)}	&\multicolumn{1}{c}{$\tilde{\nu}_{fi}$}	\\
\hline\hline
       14123   &    13911&  1.5571E-02    &    10159.167959 \\
       13337   &    13249&  5.9470E-06    &    10159.170833 \\
        1483   &     1366&  3.7119E-03    &    10159.177466 \\
        9072   &     8970&  1.1716E-04    &    10159.177993 \\
        1380   &     1469&  3.7119E-03    &    10159.178293 \\
       14057   &    13977&  1.5571E-02    &    10159.179386 \\
       10432   &    10498&  4.5779E-07    &    10159.187818 \\
       12465   &    12523&  5.4828E-03    &    10159.216008 \\
       20269   &    20286&  1.2448E-10    &    10159.227463 \\
       12393   &    12595&  5.4828E-03    &    10159.231009 \\
        2033   &     2111&  6.4408E-04    &    10159.266541 \\
       17073   &    17216&  4.0630E-03    &    10159.283484 \\
        5808   &     6085&  3.0844E-02    &    10159.298459 \\
        5905   &     5988&  3.0844E-02    &    10159.302195 \\
       13926   &    13845&  1.5597E-02    &    10159.312986 \\
\hline
\end{tabular}

\noindent
 $f$: Upper  state counting number;\\
$i$:  Lower  state counting number; \\
$A_{fi}$:  Einstein-A coefficient in s$^{-1}$; \\
$\tilde{\nu}_{fi}$: transition wavenumber in \cm.\\

\end{table}

\section{Results}	\label{results}

\subsection{Partition Function}	\label{pfn}

The partition function is given by
\begin{equation}
\label{eq:partition}
Q(T) = g_{\rm ns} \sum\limits^{n}_{i=0} (2 J_i + 1) \exp\Big(\frac{-c_2 \tilde{E}_i}{T}\Big),
\end{equation}
$\tilde{E_i}$ is the energy term value (\cm); $c_2$ is the second
radiation constant (K~cm); and $g_{\rm ns}$ is the nuclear statistical
weight. This was calculated from the new line list using the in-house
program \textsc{ExoCross} \citep{ExoCross} up to a temperature of 5000~K in increments
of 1~K. Tabulations of this form are given in the supplementary
material for all six of the isotopologues considered.

\begin{figure}
\centering
  \centering
  \includegraphics[scale=0.3]{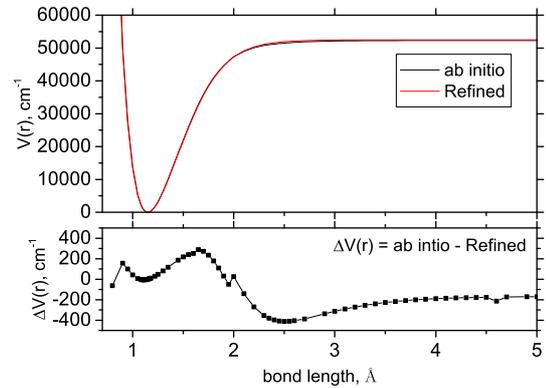}
\caption{Comparison of the calculated partition function (solid line) and that modelled by \protect\citet{84SaTaxx.partfunc} (dashed line) up to 5000~K.}
\label{fig:pfn_comp}
\end{figure}

The computed partition function compares well to the
values by \citet{84SaTaxx.partfunc}, above their lower temperature limit of 1000~K, as shown in
Fig.~\ref{fig:pfn_comp}. Slight disagreement at higher temperatures may be
due to the fact that only the ground electronic state  of NO has been
considered in the \textsc{Duo} calculations, since excited states will
have a larger contribution at high temperatures. Looking at the
log-plot comparison, disagreement below $\log(T)$ = 3.0 corresponds to
temperatures lower than 1000~K, for which the Sauval and Tatum model
is not valid (see also Table~\ref{table:pf}, where the partition functions
for temperatures are compared).

\begin{table}
\caption{Partition functions of NO: HITRAN values (TIPS) \citep{00GaKeHa.partfunc} (provided only between 70 and 3000~K),
obtained using parameters from \citet{84SaTaxx.partfunc} and \Duo\ values. }
 \begin{tabular}{rrrr}
\hline
\multicolumn{1}{c}{$T$ (K)}       & \multicolumn{1}{c}{HITRAN}       &\multicolumn{1}{c}{Sauval \& Tatum}           &\multicolumn{1}{c}{\Duo}          \\
\hline
      70&       189.75&      492.80&      193.53 \\
     100&       293.36&      585.91&      296.51 \\
     300&      1160.75&     1296.85&     1159.66 \\
    1000&      4877.73&     4874.19&     4877.53 \\
    1500&      8403.66&     8470.16&     8424.34 \\
    2000&     12812.56&    12951.71&    12887.04 \\
    2500&     18135.16&    18343.44&    18311.46 \\
    3000&     24382.24&    24673.57&    24726.00 \\
    4000&             &    40270.57&    40622.12 \\
    5000&             &    59994.10&    60769.42 \\
\hline
\end{tabular}
\label{table:pf}
\end{table}

The partition function was also represented using the following
functional form \citep{00ViTexx.H2O} given by
\begin{equation}
\label{eq:fit1d}
\log_{10}Q(T) = \sum\limits^{10}_{n=0} a_n (\log_{10}T)^n.
\end{equation}
This expression was used to least-squares fit eleven expansion
coefficients, $a_0, \ldots , a_{10}$, to the \textsc{Duo} partition
function. An example of expansion parameters for \NO{14}{16} are presented in Table
\ref{table:fit1d}. These parameters reproduce the temperature dependence of
partition function of NO within within 0.3\% for most of the data, however it
increases to just 0.4\% at $T$ = 4000~K and 1.1\% at $T$ = 5000~K.
This is still a very small error, and thus the fit can be said to
reliably reproduce the partition function. Expansion parameters for all six species are
included into the supplementary materials.

\begin{table}
\caption{Expansion coefficients for the partition function of \NO{14}{16} given by
Eq.~(\ref{eq:fit1d}). Parameters for other isotopologues can be found in the supplementary material. }
 \begin{tabular}{cc}
 \hline\hline
Expansion coefficient & 	\\
\hline\hline
$a_0    $&  1.0761409513   \\
$a_1    $& -0.1681972157   \\
$a_2    $&  1.5810964843   \\
$a_3    $& -4.5662697659   \\
$a_4    $&  9.4920289544   \\
$a_5    $& -10.9491757465  \\
$a_6    $&  7.3756190305   \\
$a_7    $& -2.9829630362   \\
$a_8    $&  0.7131937052   \\
$a_9    $& -0.0928960661   \\
$a_{10} $&   0.0050821171  \\
\hline
\end{tabular}
\label{table:fit1d}
\end{table}

\subsection{Intensities}	\label{int}

The absorption line intensities  were obtained using
\citep{05Bernath.book}
\begin{equation}
\label{eq:int}
I = \frac{1}{8\pi c \tilde{\nu}^2} \frac{g_{\rm ns}(2J' + 1)}{Q(T)} \hspace{1.5mm} A_{if} \hspace{1.5mm} \exp\Bigg({\frac{-c_2 \tilde{E}\pp}{T}}\Bigg) \Bigg[1-\exp\Bigg({\frac{c_2\tilde{\nu}}{T}}\Bigg)\Bigg] ,
\end{equation}
where $I$ is the line intensity (cm~molecule$^{-1}$); $c$ is the speed
of light (cm~$s^{-1}$); $Q(T)$ is the partition function;
$\tilde{E}\pp$ is the lower state term value; $c_2$ is the second
radiation constant (K~cm); and $g_{\rm ns}$ is the nuclear statistical
weight.

\begin{figure}
\centering
\includegraphics[scale=0.3]{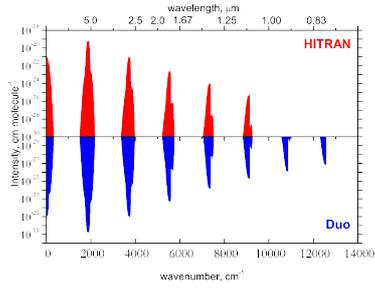}
\caption{Log-scale comparison of absorption intensities ($\mathrm{cm \hspace{1mm} molecule^{-1}}$) at $T$ = 296~K of the HITRAN database \protect\citep{HITRAN2012} (red) and this work (blue). Each intensity `column' represents a vibrational band.}
\label{fig:hitran_comp}
\end{figure}

Absorption intensities were calculated using
\textsc{ExoCross} and the lines are presented as stick spectra.
Figure~\ref{fig:hitran_comp} compares the
computed absorption intensities to intensities from
HITRAN  at 296~K \citep{HITRAN2012} up to a wavenumber of 15,000 \cm. It can be seen that the absorption intensities calculated in this work are in excellent agreement with
those of the HITRAN database, as they are of the same
strength and wavenumber; this work is more comprehensive, as
absorption intensities are calculated up to 40,000 \cm\ whilst the
HITRAN database employs a cut-off wavenumber of approximately
10,000 \cm. It should be noted that the HITRAN data is reasonably complete
at $T=$ 296~K for $^{14}$N$^{16}$O,  but not for other isotopologues (see Table~\ref{table:iso}).
HITRAN also contains a huge number of extremely weak (at 296~K) transitions (down to $10^{-95}$ cm/molecule).
Many of the strong lines are with the hyperfine structure resolved. After excluding the weak lines  (using the HITRAN cut-off algorithm \citep{HITRAN2012}) and averaging over the hyperfine components, we have obtained about 6,400 transitions ($T=296$~K). This can be compared to 8,274 lines in our \NO{14}{16}\ line list at 296~K (using the same HITRAN cut-off). This and other comparisons  are summarised in Table~\ref{table:iso}.



\begin{figure}
\centering
  \centering
  \includegraphics[scale=0.25]{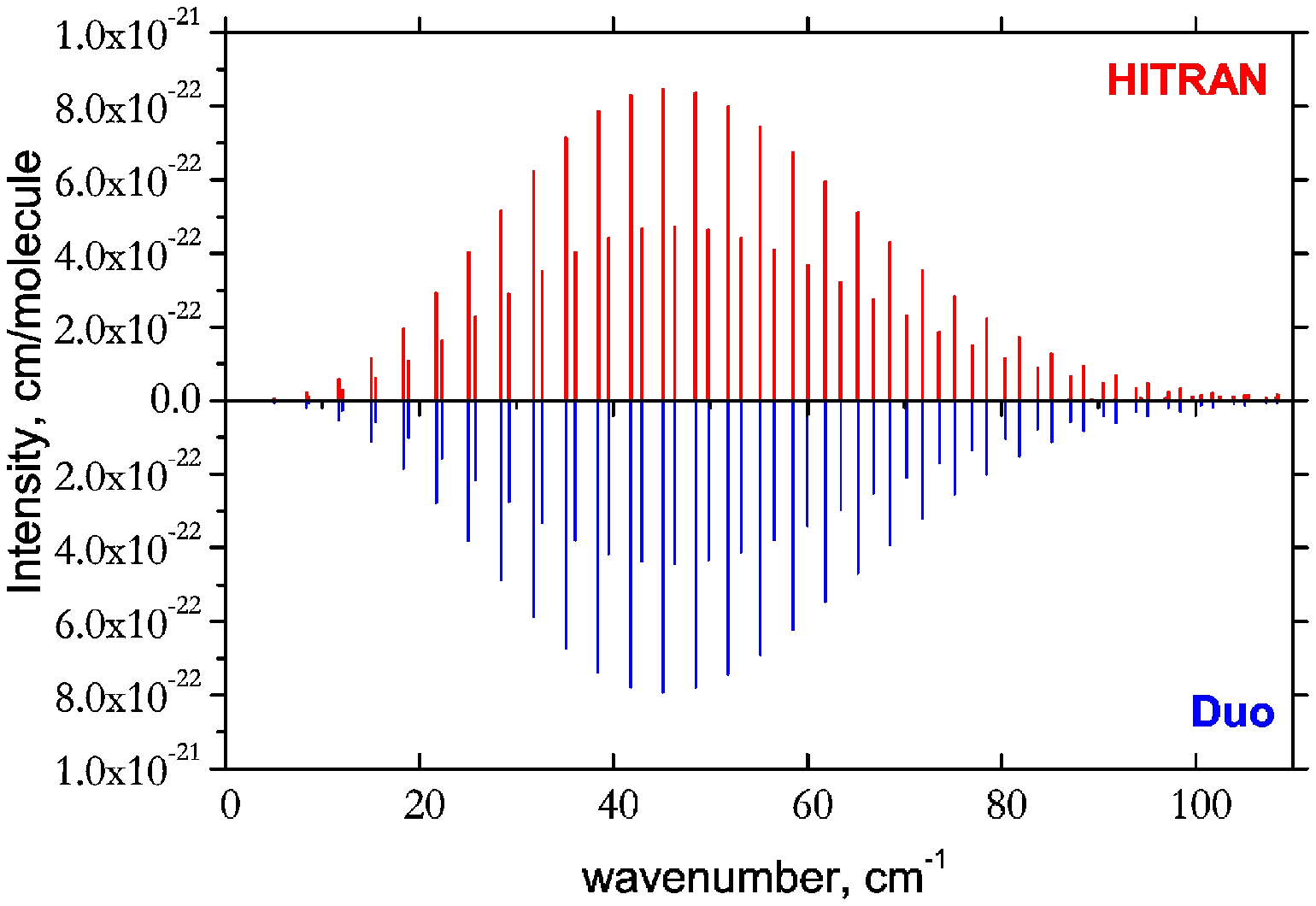}
  \label{fig:pure_rot}
  \centering
  \includegraphics[scale=0.25]{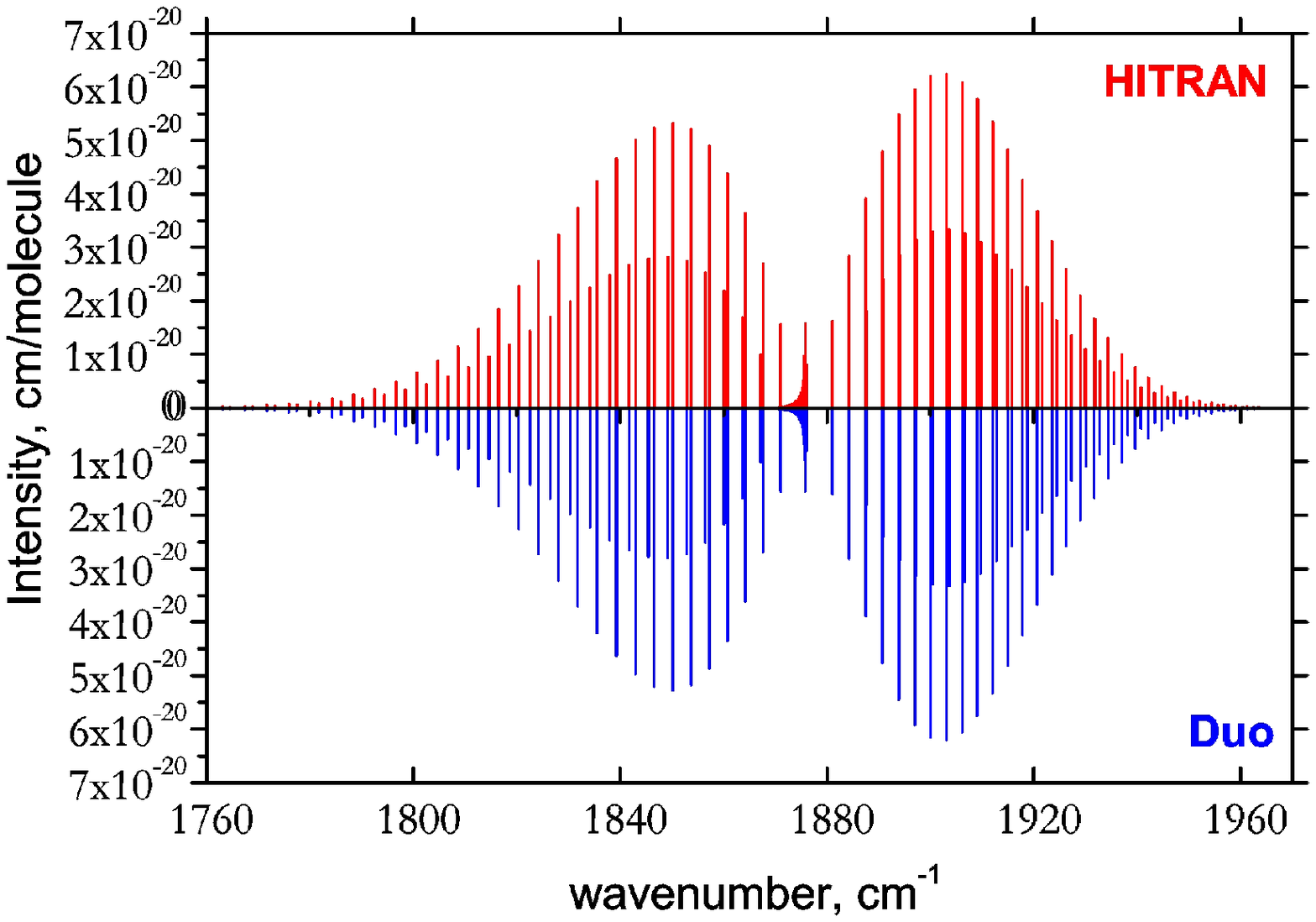}
  \label{fig:fundamental}
  \centering
  \includegraphics[scale=0.25]{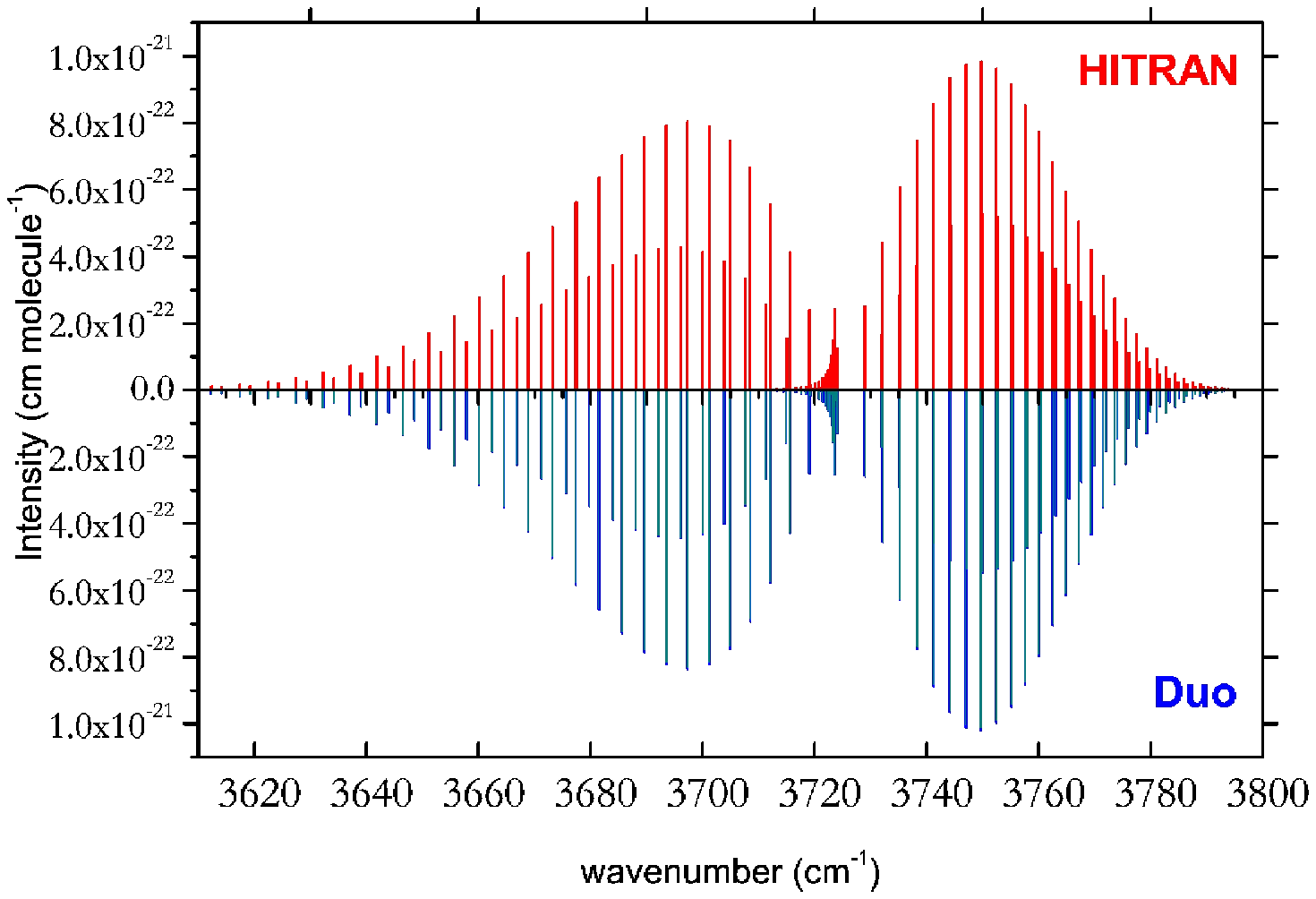}
  \label{fig:1st_overtone}
  \centering
  \includegraphics[scale=0.25]{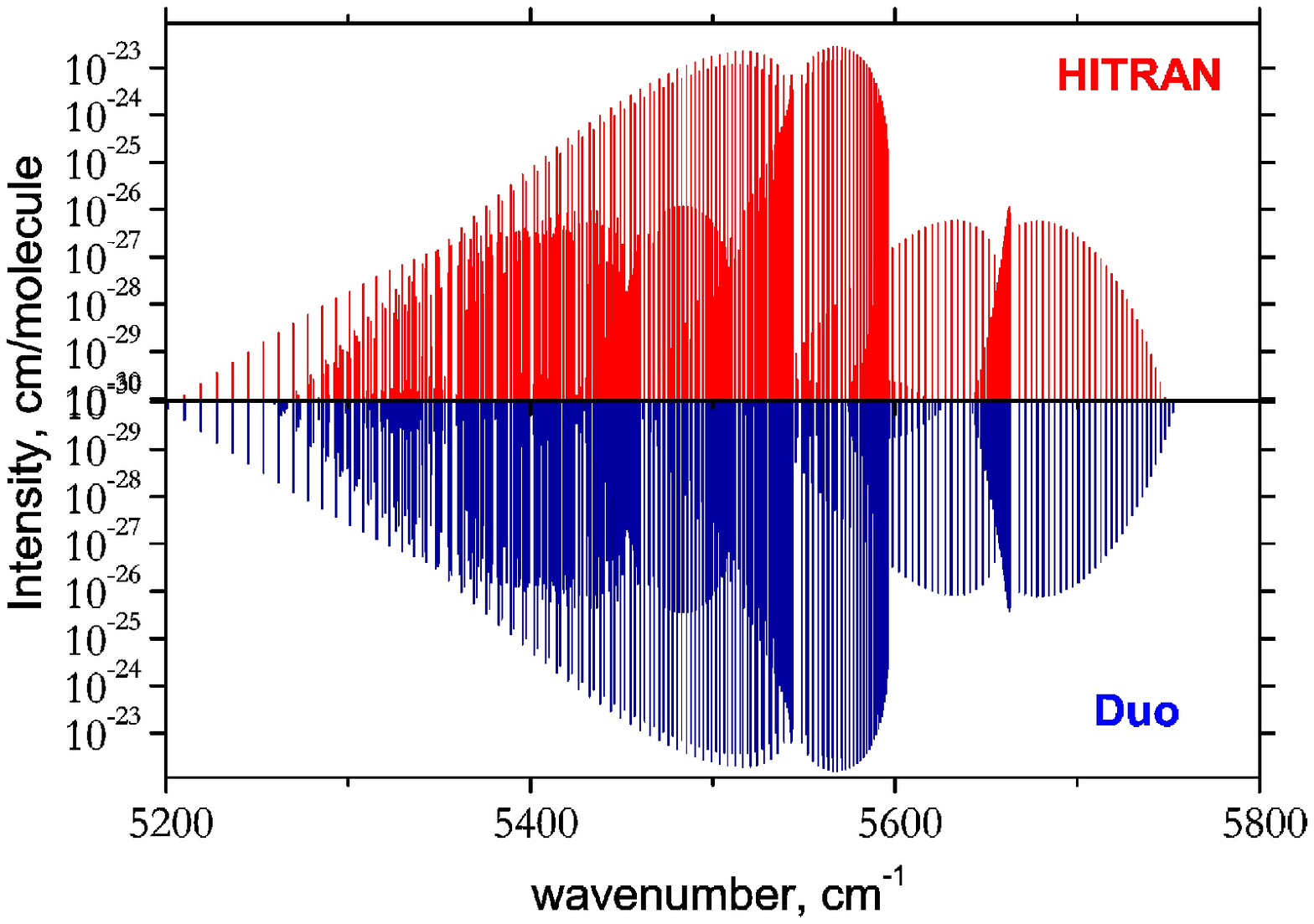}
\caption{Stick spectra comparison of HITRAN absorption intensities (cm/molecule) (red) and absorption intensities calculated in this work (blue), at a temperature of 296~K: pure rotational band, fundamental vibrational band, 1st vibrational overtone band and 2nd vibrational overtone band. Intensity strength and wavenumber positions are in excellent agreement.}
\label{fig:hitran_bands}
\end{figure}

Comparison of band structure is presented in
Fig.~\ref{fig:hitran_bands}, again comparing this work to the
HITRAN database. The pure rotational band is present in the
far-infrared region, the fundamental band is is the mid-infrared
region and the first and second vibrational overtones are present in
the near-infrared region. Branch structure is visible, with the extent
of the $P$-branch increasing with each successive overtone, whilst the
$R$-branch becomes more dense, as expected \citep{04Hollas.book}.
The fundamental band is the strongest, while the band intensity
decreases with each successive overtone as expected.

\subsubsection{Isotopologue Intensity Comparison}	\label{iso}

\begin{figure}
\centering
\includegraphics[scale=0.3]{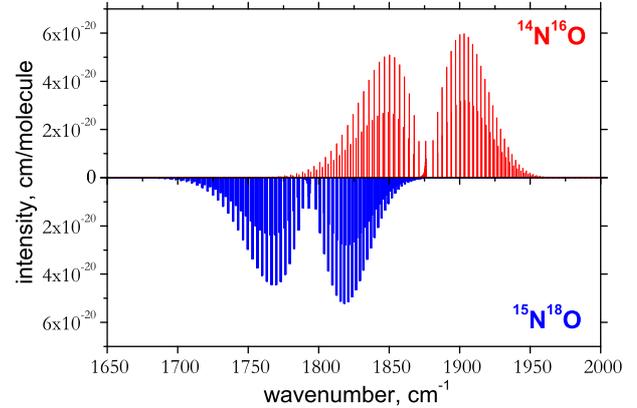}
\caption{Stick spectrum comparison of the $^{15}$N$^{18}$O fundamental vibrational band (red) and the  $^{14}$N$^{16}$O  fundamental vibrational band (blue)
in the 5.3~$\mu$m region at 296~K. Note that the $^{15}$N$^{18}$O band is shifted to lower wavenumbers, and intensities are slightly weakened.}
\label{fig:N15O18_iso}
\end{figure}

For comparison purposes, intensities were calculated using the same
procedure for the $^{15}$N$^{18}$O isotopologue; the fundamental
vibrational band is compared to the same region of the NO spectrum in
Fig.~ \ref{fig:N15O18_iso}. Since the reduced mass $\mu$ is less for
the $^{15}$N$^{18}$O isotopologue, it follows that the vibrational frequency and band origin is decreased. As a consequence, the
absorption intensities are slightly weaker, since the Einstein A
coefficients are proportional to the wavenumber cubed. This can be
seen in Fig.~\ref{fig:N15O18_iso}, as the $^{15}$N$^{18}$O band
is shifted to a lower wavenumber, and intensities are slightly
weakened.

\begin{table*}
\caption{A summary of the ExoMol isotopologue line lists (number of lines and states) and summary of
on the \textsc{SPCAT} data (number of states and $\varv_{\rm max}$). $J_{\rm max}$ = 185.5 (ExoMol) and
$J_{\rm max}$ = 99.5 (\textsc{SPCAT}). $\varv_{\rm max}$(ExoMol) = 51. `Abund' refers to terrestrial isotopic abundances.   $N_{296}$ gives
the number of NO transitions in our line lists at 296~K after  after applying the HITRAN intensity cutoff. $N_{\rm Trans.}$ are the corresponding
numbers of lines in HITRAN 2012 \citep{HITRAN2012} (neglecting hyperfine structure). }
\begin{tabular}{rlrrrrrrr}
\hline\hline
            &     & \multicolumn{3}{c}{ExoMol} & \multicolumn{2}{c}{\textsc{SPCAT}} &   \multicolumn{2}{c}{HITRAN} \\
\cline{3-9}
isotopologue & Abund & $N_{\rm States}$ & $N_{\rm Trans.}$ & $N_{296}$ & $N_{\rm States}$ & $\varv_{\rm max}$ & $N_{296}$ & $N_{\rm Trans.}$ \\
\hline
$^{14}$N$^{16}$O     &  0.995      &     21,688 &   2,281,042 &      8,274 &    11,940 &     29 &  93,622   &    6,369 \\
$^{14}$N$^{17}$O     &  0.000379   &     22,292 &   2,378,578 &      3,067 &     1,990 &      4 &           &          \\
$^{14}$N$^{18}$O     &  0.00205    &     22,848 &   2,471,705 &      3,853 &     3,980 &      9 &    679    &    679   \\
$^{15}$N$^{16}$O     &  0.00363    &     22,466 &   2,408,920 &      4,233 &     3,980 &      9 &    699    &    699   \\
$^{15}$N$^{17}$O     &  0.00000138 &     23,106 &   2,516,634 &      1,290 &     1,990 &      4 &           &          \\
$^{15}$N$^{18}$O     &  0.00000746 &     23,698 &   2,619,513 &      1,790 &     1,990 &      4 &           &          \\
\hline\hline
\end{tabular}
\label{table:iso}
\end{table*}

\subsection{Cross-Sections}	\label{xsec}

\begin{figure}
\centering
  \includegraphics[scale=0.3]{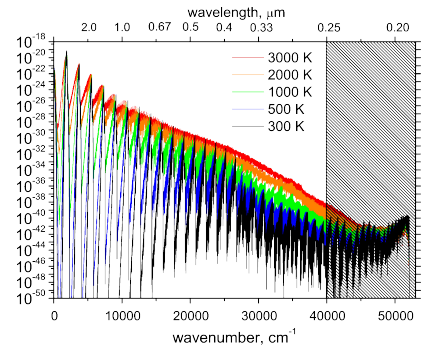}
  \caption{Absorption spectrum of the ground state of  $^{14}$N$^{16}$O  as a function of temperature.
  The temperatures considered are 300~K (bottom), 500~K, 1000~K, 2000~K and 3000~K (top). Cross-sections are calculated with a Gaussian profile and HWHM = 1 \cm. The higher-temperature profiles will be useful in characterising the spectra of terrestrial exoplanets and brown dwarfs.}
  \label{fig:abs_temps}
\end{figure}

Figure~\ref{fig:abs_temps} shows absorption cross-sections  computed at temperatures
of 300~K, 500~K, 1000~K, 2000~K and 3000~K using \textsc{ExoCross}
for the wavelength range up to 0.2~$\mu$m.
A Gaussian line profile was specified, with a half width at half maximum (HWHM) of 1~\cm.
The intensities drop with the wavenumber (overtones) exponentially as they
should~\citep{15LiGoRo.CO} up to 40000~\cm\ (the upper bound in our line lists),
after which plateau-like structures start forming at very small intensities ($<10^{-40}$ cm$/$molecule).
The latter indicates the artifacts in our dipole at very high vibrational excitations.
Transitions with wavelength less than 0.25~$\mu$m,  indicated by the shaded area in \ref{fig:abs_temps}, are are not included in the NO line lists.

\begin{figure}
\centering
\includegraphics[width=8cm]{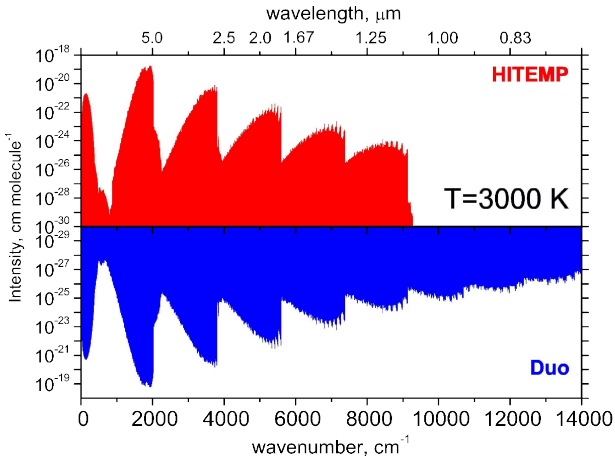}
\vspace{1cm}
\caption{Comparison of absorption cross-sections (cm$^2$/molecule) with a Doppler line profile at 3000~K of the HITEMP database \citep{jt480} (red) and this work (blue).}
\label{fig:hitemp_comp}
\end{figure}

Absorption cross-sections of \NO{14}{16} with a Doppler profile were computed from
the HITEMP database for $T$ = 3000~K, in the range 0 --
14,000 \cm\ and compared to cross-sections generated using the \NO{14}{16} ExoMol line list, see
Fig.~\ref{fig:hitemp_comp}. There is a good general agreement in strength
and wavenumber between the two spectra.  Again, it should be noted that
this work is more extensive, as the HITEMP database employs a
cut-off wavenumber of approximately 10,000 \cm, as does the
HITRAN database.

\subsection{Radiative Lifetimes}	\label{life}

\begin{figure}
\centering
\includegraphics[scale=0.3]{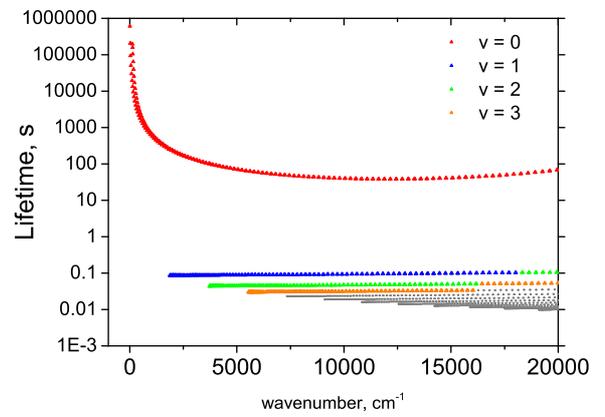}
\caption{A log-plot of  $^{14}$N$^{16}$O radiative lifetimes against state energy. Lifetimes for states with  $\varv=0-3$ are indicated by triangles while lifetimes for higher vibrational states are indicated by circles.}
\label{fig:life}
\end{figure}

The radiative lifetime of an excited state, $\tau_i$,
can be computed in a straightforward manner from the state and transition files \citep{jt624} by
\begin{equation}
\label{lifetimes}
\tau_i = \frac{1}{\sum\limits_{f<i}A_{if}}
\end{equation}
where $A_{if}$ is the Einstein A coefficient, and $i$ and $f$ indicate the
initial and final states, respectively.  Lifetimes were calculated by the program
\textsc{ExoCross}. The computed lifetimes are plotted in
Fig.~\ref{fig:life} as a function of wavenumber (\cm); lifetimes for
all states are plotted in grey, whilst lifetimes for the $\varv = 0-3$
states are highlighted by coloured triangles. Lifetimes for states
for which all downwards transitions are considered are given as
part of the enhanced ExoMol states file \citep{jt631} as illustrated in
 Table \ref{table:states} .

\section{Discussion and Conclusion}	\label{conc}

The line list called \textsc{NOname} for the ground state of the NO isotopologue
$^{14}$N$^{16}$O was constructed using a hybrid (variational/effective Hamiltonian) scheme.
The line list contains 21,688 states and 2,409,810 transitions in the wavenumber range
0--40,000 cm$^{-1}$, extending to maximum quantum numbers
${J=184.5}$ and ${v=51}$. Line lists were also constructed for the five
isotopologues, $^{14}$N$^{17}$O, $^{14}$N$^{18}$O,
$^{15}$N$^{16}$O, $^{15}$N$^{17}$O and $^{15}$N$^{18}$O in the same range and containing similar
numbers of states and transitions.

Initial energy levels in the line lists were calculated by a fit of \ai\
results using experimental energies.  Refinement of the energy levels
returned an rms of 0.015~\cm, which corresponds to a fit which is
accurate to 0.02~\cm\ for 80\% of the data, whilst the worst residual
is 0.13~\cm. These were then replaced by semi-empirical energies, where available.
The accuracy of the energy levels propagates through to the computed line lists; comparison of
intensities from this work and the HITRAN \citep{HITRAN2012} database for the $^{14}$N$^{16}$O isotopologue at 296~K show excellent agreement both in strength and position of lines. Because most of the \Duo\ energies were replaced with the semi-empirical ones, the fit was mostly done  to improve the accuracy of intensities via better quality of the corresponding wavefunctions. Only highly excited states ($J>100.5$ and $\varv>29$ for \NO{14}{16}) were taken from the \Duo\ calculations, shifted at the stitching points to avoid discontinuities. Thus our \NO{14}{16}  line positions can be considered of experimental accuracy for $\varv \le 22$, which is then expected to degrade gradually when extrapolated to $\varv = 22 \ldots 51$. The difference between \textsc{SPCAT} and \Duo\ at $\varv = 29$ (the stitching point) is 2.47~\cm, after which we rely on the \Duo\ extrapolation. It should be noted, however, that the impact from the energies in the extrapolated region is marginal for practical applications due the low absorption intensities of the corresponding transitions. For the example, the overtones with $\varv'>29$ fall into the wavenumber region above 40,000~\cm, which is fully excluded from -the line list. We keep the corresponding energies anyway for the sake of completeness.

The partition function $Q(T)$ was calculated for the
$^{14}$N$^{16}$O isotopologue, and compared to that computed by
\citet{84SaTaxx.partfunc}; there is good agreement above 1000~K, below
which the Sauval and Tatum model is not valid. Slight disagreement at
high temperatures is likely due to the fact that only the ground state
of NO is considered in this work, since excited states will have a
larger contribution to the partition function at high temperatures. An
idea for future work is to compute line lists for the excited states,
and to model the interaction between these states, in order to improve
the accuracy of the line list at high rotational and vibrational
energy levels.

Lifetimes were calculated for all energy levels considered. Absorption
cross-sections have been calculated for temperatures ranging from
300~K $-$ 5000~K. The absorption spectrum at 3000~K is in excellent
agreement with but much more extensive than the same spectrum calculated from the HITEMP
database \citep{jt480}, illustrating that the line list is also
accurate at high temperatures. The absorption spectra will be applied
in the characterisation of high-temperature astronomical objects such
as exoplanet atmospheres, brown dwarfs and cool stars. The NO spectra may
also be useful in the remote sensing of high temperature events in the Earth's atmosphere
such as lightning and vehicle re-entry from orbit. Our calculations
also provide Land\'e $g$-factors for each state; a comparison of these values with observed \citep{11IoKlKo.NO} Zeeman splitting of NO states in weak magnetic fields was carried out by \citet{jt656}, and found very good agreement.

The six NO line lists are the most comprehensive
available; they extend up to a wavenumber of 40,000~\cm, compared to
the upper limit of 10,000~\cm\ in both the HITRAN and
HITEMP databases \citep{jt480,HITRAN2012}.
These line lists can be downloaded from the CDS, via
ftp://cdsarc.u-strasbg.fr/pub/cats/J/MNRAS/, or
http://cdsarc.u-strasbg.fr/viz-bin/qcat?J/MNRAS/, or from www.exomol.com.
On the ExoMol website we also provide a script to convert the line list into the native
HITRAN format.


\label{lastpage}



\section*{Acknowledgements}

This work is supported by ERC Advanced Investigator Project 267219.
We also acknowledge the networking support by the COST Action CM1405 MOLIM.
Some support was provided by the NASA Exoplanets program.
JT and SY thank the STFC project ST/M001334/1, UCL for use of the Legion High Performance Computer and DiRAC@Darwin HPC cluster. DiRAC is the UK HPC facility for particle physics, astrophysics, and cosmology and is supported by STFC and BIS.


\bibliographystyle{mnras}

\end{document}